\documentclass[]{rAMF2e}

\usepackage{lineno}
\usepackage{bbm}
\usepackage{url}
\begin{document}



\title{A Numerical Study of Radial Basis Function Based Methods for Options Pricing under the One Dimension Jump-diffusion Model}

\author{RON T.L. CHAN and SIMON HUBBERT\thanks{{\em{Correspondence Address}}: Ron T.L. Chan and Simon Hubbert, Department of Economics, Mathematics and Statistics, Birkbeck, School of Business, Economics and Informatics, University of London, Malet Street, London WC1E 7HX, UK. Email: ronctlsmile@yahoo.co.uk/s.hubbert@bbk.ac.uk\vspace{6pt}}}

\affil{Department of Economics, Mathematics and Statistics, Birkbeck, School of Business, Economics and Informatics, University of London, Malet Street, London WC1E 7HX, UK.}

\received{v1.1 released August 2010, v1.2 released December 2010, v1.3 released October 2011}

\maketitle

\begin{abstract}
The aim of this paper is to show how option prices in the Jump-diffusion models, mainly on the Merton and Kou models,  can be computed using meshless methods based on Radial Basis Function (RBF) interpolation. The RBF technique is demonstrated by solving the partial integro-differential equation (PIDE) in one-dimension for the American vanilla put and the European vanilla call/put options on dividend-paying stocks. The radial basis function we select is the Cubic Spline. We also propose a simple numerical algorithm for finding a finite computational range of an improper integral term in the PIDE so that the accuracy of approximation of the integral can be improved. Moreover, we use a numerical technique called factorization of the Cubic Spline to avoid inverting the ill-conditioned Cubic Spline interpolant. Finally, we will show numerically that in the European case the solution is second order accurate for the spatial and time variables, while in the American case it is second order accurate for spatial variables and first order accurate for time variables.

\end{abstract}

\section{Introduction}
\label{intro}
In this paper we show how to compute European and American option prices in the Jump-diffusion model using
Radial Basis Function (RBF) interpolation techniques. RBF methods have recently been proposed for numerically solving initial value and free boundary problems for the classical Black and Scholes
equation, both in the one and in the multiple asset case \citep{Fas_Kha_Vos_2:2004,Fas_Kha_Vos_1:2004,Hon_Mao:1999,Lar_AAhl_Hall:2008}. The new feature of the present
paper is that in the Jump-diffusion model, as in general L\'evy type models, the Black and Scholes PDE is replaced by a Partial Integro-Differential Operator or PIDE, involving a global term in the form of an integral operator. The PIDE has a form:
\begin{eqnarray}\label{eqn:PIDE}
\partial _{\tau } u(x, \tau ) &=& \frac{1 }{2 }
\sigma ^2 \partial _x ^2 u + \bigg(r - q - \frac{1}{2}\sigma^2-\eta\bigg)\partial _x u -(r+\lambda)u+\nonumber \\
&\ &\lambda \int _{\mathbb{R }} u(x + y , \tau )  f(y) dy
\end{eqnarray}
\citep[cf.][]{Con_Tan:2004,Schoutens:2003,Scho:2006}. Our main contribution is to show how to numerically solve (\ref{eqn:PIDE}) in an efficient way using RBFs, both for initial value and free boundary
problems (as for American options). We have chosen the Jump-diffusion model as a typical case on which to test the present RBF methodology. Our method extends however without problems to other contexts in which the
basic pricing equation is a PIDE, like that of L\'evy-type models such as Carr-Geman-Madan-Yor (CGMY) \citep{Car_Ger_Mad_Yor:2002} or Variance Gamma (VG) \citep{Carr_Madan_Chang:1998,Mad_Mil:1991}. These will be treated in a future paper.

Currently, PIDEs such as the Merton Model \citep{Merton:1976} and the Kou Model \citep{Kou:2002,Kou_Wang:2001}, one have mostly been treated by a traditional Finite Difference Method (FDM) or Finite Elements Method (FEM). In FDM, the idea is to simply fully discretize the PIDE on an equidistant grid, after having (artificially) localized the equations to some bounded interval/domain in $\mathbb{R } $. The global integral
term can be computed by numerical quadrature or by using the Fast Fourier Transform (FFT) \citep[see,][]{Almendra:2004, Alm_Oos:2004,Alm_Oos:2005, Alm_Oos:2006, Alm_Oos_CGMY:2007, Ander_Andre:2000, Bri_Nat_Rus:2007, Con_Vol:2005, Hall_For_Lab:2004, Hall_For_Vet:2005, Hir_Mad:2004,Wang_Wan_Fory:2007}. By contrast, FEM is defined as piecewise polynomial functions or wavelet functions on regular triangularizations. This technique is used to approximate solutions of the partial differential terms as well as of the integral term \citep[cf.][]{Alm_Oos:2005, Mat_Nit_Sch:2003, Mat_Sch_Wih:2005}.

In general, there is a problem which arises with these current approaches. Some of the literature, e.g. \citep{Ander_Andre:2000, Bri_Nat_Rus:2007, Con_Vol:2005}, plays down the importance of pricing American and European vanilla option values when time to maturity is less than 3 months. The reason is that for short times-to-maturity the numerical methods used to price the option tend to be inaccurate near the strike price where a singularity
(kink) exists. A singularity is defined as a point at which the function, or its derivative, is discontinuous. The payoff functions of vanilla call and put options have such a singularity. As a result, standard numerical methods such as FDM with Crank-Nicolson and without any adaptive schemes cannot ensure accuracy of option prices around the strike and a substantial amount of oscillation occurs around the strike when Option Delta $\Delta$ and Gamma $\Gamma$ are approximated \citep{Gil_Car:2006}. Giles and Carter shed light on this kind of problem \citep{Gil_Car:2006} by suggesting Rannacher's time stepping method. This is a mixture of four half-timsteps of backward Euler and Crank-Nicolson methods. Although they solve an one dimensional PDE under the Black-Scholes model and Heston's volatility model rather than a PIDE under L\'evy models, their methods of using backwards Euler timestepping in one or more initial timesteps have been proved to be achieved second-order convergence in a European case. They also carry out a detail error analysis of their methods by using Fourier analysis and find out four half-timesteps of backward Euler time-marching is the minimum require to recover second-order convergence of solving the PDE. Forysth \textsl{et al.}\citep{Hall_For_Vet:2005} also use the similar idea by suggesting Rannacher's time stepping method \citep{Ran:1984} to solve a PIDE under the Merton Jump-diffusion model. They demonstrate this technique by approximating an option price whose maturity is a quarter of a year. This method gives second order rates of convergence when pricing European options but not American ones. By using the same idea and combining it with a penalty method and a modified form of a timestep selector suggested in \citep{John:1987}, Forysth \textsl{et al}. \citep{Hall_For_Lab:2004} show how to achieve second order convergence for pricing American options. Unfortunate they do not carry out any stability analysis when they apply Rannacher's time stepping method to solve the PIDE. Moreover there is no minimum requirement of choosing half-timsteps of backward Euler before Crank-Nicolson methods are applied. All they do is by trial and error.

In most recent research papers in quantitative finance \citep[cf.][]{Fas_Kha_Vos_2:2004, Fas_Kha_Vos_1:2004, Hon_Mao:1999, Lar_AAhl_Hall:2008} RBF-approximation methods with Multiquadric (MQ) as a basis function have been proposed for numerically solving the classical Black and Scholes PDE, both in the one and in the multiple asset case. In this literature MQ is a more favorite choice than other radial basis functions, such as Thin plate Spline and the like, because of its comparatively higher accuracy. MQ contains a shape parameter which plays an imperative role in the accuracy of the method \citep[cf.][]{Wendland:2005}. Most of this recent literature still chooses this parameter by trial and error or some other ad-hoc means. Although there exists a substantial literature on choosing an "optimal" shape parameter in MQ, e.g. \citep{Fas_Zha:2007}, \citep{Forn_Wrig:2004} and \citep{Kan_Car:1992}, it is still an open question and there is no theoretical proof for selecting an optimal shape parameter \citep[cf.][]{Wendland:2005} in MQ. Besides this, the standard approach to the solution of the radial basis function interpolation problem has been recognized as an ill-conditioned problem for many years \citep[cf.][chapter 16]{Fas:2007}. This is especially true when infinitely smooth basic functions such as MQ or Guassian are used with small values of their associated shape parameters. More recently, Fasshauer and Mccourt's least-squares approximation based on early truncation of the kernel expansion \citep{Fas_Mcc:2010}and Fornberg and co-workers's Contour-Pad\'e integration method \citep[e.g.][]{Dri_For:2002, Forn_Wrig:2004,  Lar_For:2005} are successful in solving the ill-conditioning problem of RBF, but the techniques are only restricted to solve the simple interpolation problem rather than to solve PDEs, especially parabolic PDEs. Although Ling and his co-workers \citep[e.g.][]{Lin_Kan:2004,Bro_Lin_Kan_Lev:2005,Lin_Kan:2005} address the ill-conditioning problem by using preconditioning methods and extend them to solve PDEs, the methods are not possible to be applied to solving PIDEs.

Our RBF-approximation method with the Cubic Spline as a basis function will circumvent these disadvantages. This paper is divided into five sections, including this introduction. Section \ref{Chptr4:section:PIDEJumpD} is a brief review of both the Merton and Kou Jump-diffusion models. In section \ref{Chptr4:section:Meshfree} we first explain and then define our RBF algorithm for solving PIDEs, which we implement the Jump-diffusion model. Section \ref{Chptr4:section:results} contains our numerical results for both European and American call and put options, including an analysis of the max error, the root-mean-square error, the rate of convergence and the approximation of $\Delta$ and $\Gamma$ and also a comparison the accuracy of our solution with that of FDM and FEM . Section \ref{Chptr4:section:conclusions} concludes.

\section{\bf PIDE Option Pricing Formula in Jump-diffusion Market}\label{Chptr4:section:PIDEJumpD}

In this short section we will focus on the Merton and the Kou Jump-diffusion Models which are general L\'evy processes consisting of Brownian motion and compound Possion jumps. By using these models we can describe the price dynamics of the underlying risky asset, $(S_t)_{t\geq 0}$. The evolution of $(S_t)_{t\geq 0}$ is driven by a diffusion process, punctuated by jumps which describe rare events such as crashes and/or drawdowns at random intervals. As a market model, it is an example of an incomplete market.

The stock price process, $(S_t)_{t\geq 0}$, driven by these models, is given by:
\begin{align}
S_t=S_0e^{L_t}\label{Chptr4:eqn:price_process}
\end{align}
where $S_0$ is the stock price at time zero and $L_t$ is defined by:
\begin{equation}
L_t:=\gamma_c t+\sigma
W_t+\sum_{i=1}^{N_t}Y_i,\label{Chptr4:eqn:Jump_diffusion_component}
\end{equation}
here, $\gamma_c$ is a drift term, $\sigma$ is a volatility, $W_t$ is a Brownian motion, $N_t$ is a Possion process with intensity $\lambda$, $Y_i$ is an i.i.d. sequence of random variables. Since $\sigma > 0 $ in (\ref{Chptr4:eqn:Jump_diffusion_component}), there exists a risk-neutral probability measure $\mathbb{Q}$ such that the discounted process $\{e^{-(r-q)}S_t\}_{t\geq 0}$ becomes a martingale \citep[cf.][Theorems 33.1 and 33.2]{Sato:1999}, where $r$ is the interest rate and $q$ is the dividend rate. For a discussion of the issue of choosing $\mathbb{Q}$ see, for example, \citep{Con_Tan:2004}. Then under this new measure $\mathbb{Q}$, the risk-neutral L\'evy triplet of $L_t$ can be described as follows:
$$(\gamma_c, \sigma, \nu )$$
where
\begin{eqnarray}\label{Chptr4:eqn:gamma_c}
\gamma_c&=& r - q - \frac{1}{2}\sigma^2-\lambda\eta+\int _{\mathbb{R } } \, x \, \nu(dx),
\end{eqnarray}
Here we focus on the case where the L\'evy measure is associated
to the pure-jump component and hence the L\'evy measure $\nu (dx)$ can be written as $\lambda f(x)dx$, where the weight function $f(x)$ can take two forms:
\begin{enumerate}
\item In the classical Merton model, for any $i\in\{1,2,\ldots\}$, $Y_i$ are log-normally distributed variables with $Y_i\sim\mathbb{N}(\mu_J,\sigma_J^2)$ and as a result,
\begin{eqnarray}
f(x):=\frac{1}{\sqrt{2\pi}\sigma_j}e^{(x-\mu_J)^2/2\sigma_J^2}.\label{Chptr4:eqn:merton_density_function}
\end{eqnarray}
\item In the Kou model,
\begin{eqnarray}
f(x)=p\alpha_1e^{-\alpha_1x}\mathbbm{1}_{x\geq
0}+(1-p)\alpha_2e^{\alpha_2x}\mathbbm{1}_{x\leq 0}.\label{Chptr4:eqn:Kou_density_function}
\end{eqnarray}
\end{enumerate}
\begin{remark}
\rm{ In the Merton Jump-diffusion model, one should notice that $Y_i$ is i.i.d so for each $i\in\{1,2,3,\ldots\}$, $Y_i$ has the same mean and variance. For the sake of simplicity, we use $\mu_J$ and $\sigma_J^2$ to represent the mean and variance of each $Y_i$ respectively.}
\end{remark}

Also in (\ref{Chptr4:eqn:gamma_c}), $\eta=\int _{\mathbb{R } } \, (e^x - 1)f(x)\,dx$ represents the expected relative price change due to a jump. Since we have defined the L\'evy density function $f(x)$ for both Jump-diffusion processes, $\eta$ can be computed as:
\begin{enumerate}
\item In the Merton model,
\begin{eqnarray}\label{Chptr4:eqn:Merton_eta}
\eta=e^{\mu_J+\sigma_J^2/2}-1.
\end{eqnarray}
\item In the Kou model,
\begin{eqnarray}\label{Chptr4:eqn:Kou_eta}
\eta=\frac{p\alpha_1}{\alpha_1-1}+\frac{(1-p)\alpha_2}{\alpha_2+1}-1.
\end{eqnarray}
This is found by integrating $e^x$
over the real line by setting $\alpha_1>1$ and $\alpha_2>0$.
\end{enumerate}
For the details of the computation of (\ref{Chptr4:eqn:Merton_eta}) and (\ref{Chptr4:eqn:Kou_eta}), we shall refer the reader to \citep{Con_Tan:2004,Boy_Lev:2002}.

The drift-term $\gamma_c$ in (\ref{Chptr4:eqn:Jump_diffusion_component}) assumes that $e^{-(r-q)t}S_t$ is a martingale with respect to the natural filtration. We let $\tau=T-t$, the
time-to-maturity, where $T$ is the maturity of the financial option
under consideration and we introduce $x=\log S_t$, the underlying
asset's log-price. If $u(x,\tau)$ denotes the values of some
(American and European) contingent claim on $S_t$ when $\log S_t=x$
and $\tau=T-t$, then it is well-known, see for example, \citep{Con_Tan:2004}
that $u$ satisfies the following PIDE in the non-exercise region:

\begin{eqnarray}\label{Chptr4:eqn:PIDE_MJ2}
\partial _{\tau } u(x, \tau ) &=& \frac{1 }{2 }
\sigma ^2 \partial _x ^2 u + \bigg(r - q - \frac{1}{2}\sigma^2-\eta\bigg)\partial _x u -(r+\lambda)u+\nonumber \\
&\ &\lambda \int _{\mathbb{R } } u(x + y , \tau )  f(y) dy ,\\
&=:& \mathcal{L }[u] (x, \tau ) . \nonumber
\end{eqnarray}
with initial value
\begin{eqnarray}
    u(x, 0 ) = g(x) := G(e^x )=\begin{cases}\max\{e^x-K,0\}\,\hbox{, call option} \\
    \max\{K-e^x,0\}\,\hbox{, put option}\end{cases}: \label{Chptr4:eqn:payoffCallPut}
\end{eqnarray}

For an American put, we have to take into account the possibility of
early exercise \citep[e.g.,][]{Con_Tan:2004,Schoutens:2003,Scho:2006}. As a result, the
highest value of American option can be achieved by maximizing over
all allowed exercise strategies:
\begin{equation}
u(x,\tau)=\rm{ess\,sup_{\tau^*\in
\Gamma(t,T)}}E_t^{Q}\left[e^{-r(\tau^*-t)}G\big(e^{x_{\tau^*}}\big)\right]
\end{equation}
where $\Gamma(t,T)$ denotes the set of non-anticipating exercise
times $\tau^*$, satisfying $t\leq \tau^* \leq T$. To actually
compute the $u(x,\tau)$ of the American put, one can solve the
following linear complementarity problem \citep{Con_Tan:2004,Schoutens:2003,Scho:2006}:
\begin{align}
\partial_{\tau} u(\tau,x)-\mathcal{L}u(x,\tau)&\geq 0, \,\hbox{in}\,(0,T)\times \mathbb{R}\\
u(x,\tau)-G(e^{x})&\geq 0,\,\rm{a.e.}\,\hbox{in}\,(0,T)\times \mathbb{R}\\
\big(u(x,\tau)-G(e^{x})\big)\left(\partial_{\tau} u(\tau,x)-\mathcal{L}u(x,\tau)\right)&= 0,\,\hbox{in}\,(0,T)\times \mathbb{R}\\
u(x,0)&=G(e^{x}),
\end{align}
Since we only deal with a jump-diffusion model with $\sigma>0$ and
finite jump intensity in this paper, we know that by Pham
\citep{Pham:1997}, the smooth pasting condition,
$$\frac{\partial u(x_{\tau^*},\tau^{*})}{\partial x}=-1$$ is valid
at time of exercise $\tau*$. Therefore the value of an American put
option is continuously differentiable with respect to the underlying
on $(0,T)\times \mathbb{R}$; in particular the derivative is
continuous across the exercise boundary.

\begin{remark}
One should notice that if we set $\lambda=0, $ (\ref{Chptr4:eqn:PIDE_MJ2}) will become original Black-Scholes PDE.
\end{remark}

\section{\bf Meshfree Numerical Approximation Method}\label{Chptr4:section:Meshfree}
Meshfree radial basis function (RBF) interpolation is a well-known technique for reconstructing an
unknown function from scattered data. It has numerous applications
in different fields, such as terrain modeling in geology, surface
reconstruction in imaging, and the numerical solution of partial
differential equations in applied mathematics. In particular, RBFs
have recently been used to solve the PDEs of quantitative finance. A
number of authors, including Fausshauer \textit{et al.}
\citep{Fas_Kha_Vos_2:2004, Fas_Kha_Vos_1:2004}, Larsson \textit{et al.} \citep{Lar_AAhl_Hall:2008}, Pettersson \textit{et al.} \citep{Pet_Lar_Mar_Per:2008} and Hon and Mao
\citep{Hon_Mao:1999}, have suggested RBFs as a tool for solving
Black-Scholes equations for European as well as American options.
This numerical scheme for the estimation of partial derivatives
using RBFs was originally proposed by Kansa \citep{Kan_1:1990},
resulting in a new method for solving partial differential equations
\citep{Kan_2:1990}. The aim here is to obtain a RBF approximation of the
initial value or pay-off of the option. Once we are disposition of such an
RBF-interpolant, we implement an RBF-scheme to solve
the PIDE with this RBF-interpolant as initial value. The general
idea of the proposed numerical scheme is to approximate the unknown
function $u(x,\tau)$ by an RBF-interpolant using the interpolation
points found for the initial value using the RBF-scheme, and derive
a system of linear constant coefficient ODE by requiring that the
PIDE (\ref{Chptr4:eqn:PIDE_MJ2}) be satisfied in the chosen
RBF-interpolation points.
After picking interpolation points $x_j\in \mathbb{R}$, we
approximate, for any fixed time-to-maturity $\tau$, the solution
$u(x,\tau)$ in (\ref{Chptr4:eqn:PIDE_MJ2}) by its RBF-interpolant:
\begin{eqnarray}
    u(x,\tau)\simeq\sum_{j=1}^N \rho_j(\tau)\phi(||x-x_j||_2)=:\widetilde{u}(x,\tau), \label{Chptr4:eqn:RBF}
\end{eqnarray}
Since the radial basis function does not depend on time, the time
derivative of $\widetilde{u}(x,\tau)$ in equation (\ref{Chptr4:eqn:PIDE_MJ2}) is
simply:
\begin{align}
    \frac{\partial{\widetilde{u}}(x,\tau)}{\partial{\tau}}&=\sum_{j=1}^N
    \frac{d{\rho_j(\tau)}}{d{\tau}}\phi(|x-x_j|),
    \label{dt_RBF}
\end{align}
Moreover, the first and second partial derivatives of
$\widetilde{u}(x,\tau)$ with respect to $x$ are
\begin{align}
    \frac{\partial{\widetilde{u}(x,\tau)}}{\partial{x}}&=\sum_{j=1}^N
    \rho_j(\tau)\frac{\partial{\phi(|x-x_j|)}}{\partial{x}}\label{Chptr4:eqn:D_RBF},\\
    \frac{\partial^2{\widetilde{u}(x,\tau)}}{\partial{x}^2}&=\sum_{j=1}^N \rho_j(\tau)\frac{\partial^2{\phi(|x-x_j|)}}{\partial{x}^2},\label{Chptr4:eqn:D2_RBF}
     \intertext{where for the particular case when $\phi$ is the Cubic Spline,}
     \frac{\partial{\phi(|x-x_j|)}}{\partial{x}}&=\begin{cases}\,\,\,3(|x-x_j|)^2 \quad \text{if }x-x_j>0,\\-3(|x-x_j|)^2 \quad \text{if } x-x_j<0,\end{cases}\label{Chptr4:eqn:D_RBF_cubic}\\
    \frac{\partial^2{\phi(|x-x_j|)}}{\partial{x}^2}&=6(|x-x_j|).\label{Chptr4:eqn:D2_RBF_cubic}
\end{align}
In this research we choose the Cubic Spline rather than the most popular ones, MQ and IMQ as a basis function because of its simplicity and accuracy and without containing any shape parameters.

\subsection{Transforming PIDE to A System of ODEs by RBF}
\label{Chptr4:subsection:pide_to_ode_rbf} Given a set of interpolation points
$x_1,\ldots,x_j,\ldots,x_N$ in $\mathbb{R}$, and an RBF $\phi$, we can construct
$N\times N$ matrices $\pmb{A}$, $\pmb{A}_{x}$ and $\pmb{A}_{xx}$
defined by $\big(\phi(|x_i-x_j|)\big)_{1\leq i,j\leq N}$,
$\big(\phi^{'}(|x_i-x_j|)\big)_{1\leq i,j\leq N}$ and
$\big(\phi^{''}(|x_i-x_j|)\big)_{1\leq i,j\leq N}$ respectively.
Note in case the $x_j$'s are chosen according to the Equally Spacing Method, ESM, used in
\citep{Fas_Kha_Vos_2:2004,Fas_Kha_Vos_1:2004,Hon_Mao:1999}. In brief, Equally Spacing Method is the way to choose equally spaced points in a finite interval. In the ESM, we
determine an interval $[x_{\rm{min}}, x_{\rm{max}}]$ outside of
which we can neglect the contribution of $u(x,\tau)$ to the
global integral term of a PIDE (\ref{Chptr4:eqn:PIDE_MJ2}), and
for given $N=0,1,2,\ldots,$ simply put
\begin{equation}
x_j:=x_j^{\Delta x}=x_{\min}+j\Delta x, \,j=0,1,2,\ldots,N-1
\label{Chptr4:eqn:ems_pt_chosen}
\end{equation}
where $\Delta x=(x_{\max}-x_{\min})/(N-1)$.
We also define a matrix-valued function $y\rightarrow\pmb{A}(y)$ by $\big(\phi(|x_i+y-x_j|)\big)_{1\leq
i,j\leq N}$. If we substitute $\widetilde{u}(x,\tau)$ for $u(x,\tau)$
in (\ref{Chptr4:eqn:PIDE_MJ2}) and require the PIDE to be satisfied in the
interpolation points $x_j$, we arrive at the following system of
ODEs for the vector
$\pmb{\rho}(\tau):=\big(\rho_1(\tau),\ldots,\rho_N(\tau)\big)$
\begin{eqnarray}\label{Chptr4:eqn:ODE_In}
    \pmb{A}\pmb{\rho}_\tau&=&\frac{\sigma^2}{2}\pmb{A}_{xx}\pmb{\rho}+\left(r-q-\frac{\sigma^2}{2}-\lambda\eta\right)\pmb{A}_{x}\pmb{\rho}+(r+\lambda)\pmb{A}\pmb{\rho}+\nonumber\\
    &\ &\lambda\left(\int_{-\infty}^{\infty}\pmb{A}(y)f(y)\,\mathrm{d}y\right)\pmb{\rho},
\end{eqnarray}
where $\rho_{\tau}:=\frac{\partial{\rho}}{\partial{\tau}}$, and
where we recall that $f(y)$ is the probability density of the jump
$Y_i\sim\mathbb{N}(\mu_J,\sigma_J^2):$
$f(y)=({\sigma_J\sqrt{2\pi}})^{-1}\exp\big(-(y-\mu_J)^2/2\sigma_J^2\big)$ in the Merton model, or
$f(y)=p\alpha_1e^{-\alpha_1x}\mathbbm{1}_{x\geq
0}+(1-p)\alpha_2e^{\alpha_2x}\mathbbm{1}_{x\leq 0}$ in the Kou model.
Before applying a suitable numerical integration algorithm to the
integral terms in (\ref{Chptr4:eqn:ODE_In}), we truncate the integrals from an
infinite computational range to a finite one. Briani \textsl{et
al.} \citep{Bri_Nat_Rus:2007}, Cont and Voltchkova \citep{Con_Vol:2005},
Tankov and Voltchkova \citep{Tan_Vol:2009} and d'Halluin \textsl{et al.}
\citep{Hall_For_Lab:2004, Hall_For_Vet:2005} have provided different numerical
techniques to find out a finite computational range so as to reduce
the numerical approximation errors when doing this truncation. In this thesis we shall adopt the Briani \textsl{et
al.} numerical technique to truncate the integral domain of our PIDE (cf. \citep{Bri_Nat_Rus:2007}) in both the Merton and Kou model. See \ref{AppendixA}
for a proof. Supposed $\epsilon>0$, a formula of selecting a bounded interval $[y_{-\epsilon}, y_\epsilon]$ for the set of points $y$ in the Merton case is:
\begin{eqnarray}\label{Chptr4:eqn:finite_rng_MJ}
y_\epsilon&=&\sqrt{-2\sigma_J^2\log(\epsilon\sigma_J\sqrt{2\pi}/2)}+\mu_J,\,\, \forall \, y\geq0\\
y_{-\epsilon}&=&-y_\epsilon,\,\,\forall \, y<0.
\end{eqnarray}
In the Kou model we have
\begin{eqnarray}\label{Chptr4:eqn:finite_rng_Kou}
y_\epsilon&=&\log\big(\epsilon/p\big)/(1-\alpha_1),\,\,\forall \, y\geq0\\
y_{-\epsilon}&=&-\log\big(\epsilon/(1-p)\big)/(1-\alpha_2),\,\,\forall \, y<0,
\end{eqnarray}
We therefore transform equation
(\ref{Chptr4:eqn:ODE_In}) into
\begin{eqnarray}\label{Chptr4:eqn:ODE_In_trunc}
    \pmb{A}\pmb{\rho}_\tau&=&\frac{\sigma^2}{2}\pmb{A}_{xx}\pmb{\rho}+\left(r-q-\frac{\sigma^2}{2}-\lambda\eta\right)\pmb{A}_{x}\pmb{\rho}+(r+\lambda)\pmb{A}\pmb{\rho}+\nonumber\\
    &\ &\lambda\left(\int_{y_{-\epsilon}}^{y_\epsilon}\pmb{A}(y)f(y)\,\mathrm{d}y\right)\pmb{\rho}.
\end{eqnarray}
We use matlab's adaptive Gauss-Kronrod quadrature to
evaluate the matrix of the integrals in (\ref{Chptr4:eqn:ODE_In_trunc}): this
amounts to approximating
\begin{align}
    \int_{y_{-\epsilon}}^{y_\epsilon}\phi(|x_i+y-x_j|)f(y)\,\mathrm{d}y\approx\sum_{k=1}^mw_k\phi(|x_i+y_k-x_j|)f(y_k),\label{Chptr4:eqn:AdpLQ}
\end{align}
where $w_k$ and $y_k$ are suitable quadrature weights and quadrature
points; cf. \citep{Shampine:2008} for details. To simplify notations, we
set
$$F(x_i-x_j)=\sum_{k=1}^mw_k\phi(|x_i+y_k-x_j|)f(y_k).$$
Then the integrals in equation (\ref{Chptr4:eqn:ODE_In_trunc}) will be
approximated by
\begin{align}\label{Chptr4:eqn:tramsform_integral}
    \int_{y_{-\epsilon}}^{y_\epsilon} \pmb{A}(y)f(y)\,\mathrm{d}y&\approx
    \begin{bmatrix}F(x_1-x_1)&F(x_1-x_2)&\dots&F(x_1-x_N)\\
    F(x_2-x_1)&F(x_2-x_2)&\dots&F(x_2-x_N)\\
    \hdotsfor[2.0]{4}\\
    F(x_N-x_1)&F(x_N-x_2)&\dots&F(x_N-x_N)\\
    \end{bmatrix}\nonumber\\
    &=\pmb{C}(y).
\end{align}
Substituting (\ref{Chptr4:eqn:tramsform_integral}) into equation
(\ref{Chptr4:eqn:ODE_In_trunc}), we arrive at the new approximate equation:
\begin{align}
    \pmb{A}\pmb{\rho_\tau}&=\frac{\sigma^2}{2}\pmb{A}_{xx}\pmb{\rho}+\left(r-q-\frac{\sigma^2}{2}-\lambda\eta\right)\pmb{A}_{x}\pmb{\rho}
    +(r+\lambda)\pmb{A}\pmb{\rho}+\lambda\pmb{C}(y)\pmb{\rho}.\label{Chptr4:eqn:ODE}
\end{align}
As we have known the Cubic Spline is strictly conditionally positive definite function of order 2, the invertibility of $\pmb{A}$ is not assumed without adding a real-valued polynomial of degree at most 1 in (\ref{Chptr4:eqn:RBF}) \citep[cf.][]{Wendland:2005}. Nevertheless, Bos and Salkauskas proved that $\pmb{A}$ is non-singular in a univariate case \citep[cf.][Theorem 5.1]{Bos_Sal:1987}. As a result, the invertibility of $\pmb{A}$ is still guaranteed.

Although the invertibility of $\pmb{A}$ is able to be shown for all $\phi$ of the interest, the inverse of $\pmb{A},$ $\pmb{A}^{-1}, $ may often be very ill-conditioned to solve when its size increases \citep[cf.][chapter 16]{Fas:2007}. As a result, it may be impossible to solve accurately using standard floating point arithmetic. To address this problem, we factorise $\pmb{A}$ into the following form \citep[cf.][Theorem 3.7]{Bos_Sal:1987}:
\begin{eqnarray}\label{Chptr4:eqn:factorisation_of_CubicSpline}
     \pmb{A}=\pmb{F}\pmb{C}\pmb{F}.
\end{eqnarray}
Here $\pmb{F}$ is a $N\times N$ matrix,
\begin{eqnarray}
    \begin{bmatrix}|x_1-x_1|&|x_1-x_2|&|x_1-x_3|&\dots&|x_1-x_N|\\
    |x_2-x_1|&|x_2-x_2|&|x_2-x_3|&\dots&|x_2-x_N|\\
    \vdots&\, &\ddots&\,&\vdots\\
    |x_N-x_1|&|x_N-x_2|&|x_N-x_3|&\dots&|x_N-x_N|\\
    \end{bmatrix},
\end{eqnarray}
and $\pmb{C}$ is a near tridiagonal $N\times N$ matrix,
\begin{eqnarray}
    \begin{bmatrix}h-S&\frac{h}{2}&0&\cdots&0&\frac{S}{2}\\
    \frac{h}{2}&2h&\frac{h}{2}&0&\cdots&0\\
    0&\frac{h}{2}&2h&\frac{h}{2}&\cdots&0\\
    \vdots&\, &\, &\ddots&\,&\vdots\\
    0&0&\cdots&\frac{h}{2}&2h&\frac{h}{2}\\
    \frac{S}{2}&0&\ldots&0&\frac{h}{2}&h-S\\
    \end{bmatrix},
\end{eqnarray}
where $h$ is the distance between $x_{i+1}$ and $x_i$ for $1\leq i\leq N-1$ and $S=Nh$. We also have an explicit form of $\pmb{F}^{-1}$ \citep[cf.][Lemma 3.6]{Bos_Sal:1987} which is equal to
\begin{eqnarray}
    \begin{bmatrix}\frac{h-S}{2hS}&\frac{1}{2h}&0&\cdots&0&\frac{1}{2S}\\
    \frac{1}{2h}&-\frac{1}{h}&\frac{1}{2h}&0&\cdots&0\\
    0&\frac{1}{2h}&-\frac{1}{h}&\frac{1}{2h}&\cdots&0\\
    \vdots&\, &\, &\ddots&\,&\vdots\\
    0&0&\cdots&\frac{1}{2h}&-\frac{1}{h}&\frac{1}{2h}\\
    \frac{1}{2S}&0&\ldots&0&\frac{1}{2h}&\frac{h-S}{2hS}\\
    \end{bmatrix}.
\end{eqnarray}

We perform Gaussian elimination with partial pivoting to calculate $\pmb{C}^{-1}$. Then, we multiply both sides of
(\ref{Chptr4:eqn:ODE}) by $\pmb{C}^{-1}$ and $\pmb{F}^{-1}$ and we finally obtain the
following homogeneous system of ODEs with constant coefficients:
\begin{align}\label{Chptr4:eqn:simplifed_ODE1}
    \pmb{\rho}_\tau&=\pmb{F^{-1}C^{-1}F^{-1}}\bigg(\frac{\sigma^2}{2}\pmb{A}_{xx}+\big(r-q-\frac{\sigma^2}{2}-\lambda\eta\big)\pmb{A}_{x}+(r+\lambda)\pmb{A}+\lambda\pmb{C}(y)\bigg)\pmb{\rho}\nonumber\\
    &\equiv\pmb{\Theta}\pmb{\rho}
\end{align}
where $\pmb{\Theta}$ is defined by the left hand side. After some
numerical experimentation, we found that the matrix $\pmb{\Theta}$
is very stiff. To explain why $\pmb{\Theta}$ is stiff, we shall use the following example to illustrate it. Suppose we select our maximum and minimum logarithm price $x_{\min}$
$\big(\log(S_{\min})\big)$ and $x_{\max}$ $\big(\log(S_{\max})\big)$
in (\ref{Chptr4:eqn:ems_pt_chosen}) equal to $-10$ and $10$ respectively,
then we use (\ref{Chptr4:eqn:ems_pt_chosen}) to generate a list of 100 interpolation points.
Based on the procedures and the ideas we have mentioned above we can get a $100\times100$ matrix $\pmb{\Theta}$
in (\ref{Chptr4:eqn:simplifed_ODE1}). Then we measure the stiffness ratio of $\pmb{\Theta}$. The stiffness ratio is  the quotient of the largest and the smallest eignvalues of the Jacobian matrix  $\pmb{\Theta}.$ The ratio we have is $1.2864\times10^{5}.$ This implies that (\ref{Chptr4:eqn:simplifed_ODE1}) is a stiff ODE and therefore we have to solve the ODEs by an implicit method, e.g. backward differentiation formulas (BDFs), a modified Rosenbrock formula of order 2, the trapezoidal rule or TR-BDF2, an implicit Runge-Kutta formula with a
first stage that is a trapezoidal rule step and a second stage that
is a backward differentiation formula of order two. In this paper we
use former one.

\section{\bf Numerical Results}\label{Chptr4:section:results}

\subsection{European Vanilla Options}\label{Chptr4:subsection:results:Euro_vanilla}
In this section we first present a simple scheme to construct our computational range. We then present the numerical results of our Cubic Spline approximation scheme and compare these with  Black-Scholes, Merton and Kou's analytical option price formula for both puts and calls. Beside this, we also compare the results of our Cubic Spline approximation scheme with those of the Briani \textsl{et al.} finite difference method (FD) with implicit and explicit (IMEX) scheme in \citep{Bri_Nat_Rus:2007} and the Almendral \textsl{et al.} finite element method (FE) with backward differentiation formulas of order two (BDF2) and FD with BDF2 in \citep{Alm_Oos:2005}.

\begin{figure}[htbp]
\centering
\epsfig{file=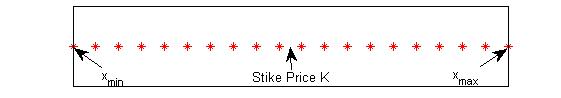}
\rule{30em}{0.5pt}
\caption{Uniform distributions of the interpolation points around the strike price by using EMS. The red dots are the interpolation points. The blue cross is the location of the logarithm strike price.
}\label{Chptr4:fig:ems_around_strike}
\end{figure}

We use EMS (\ref{Chptr4:eqn:ems_pt_chosen}) to choose our interpolation points. Based on this set of interpolation points, we can construct our computational range. We distribute the interpolation point uniformly around the logarithm strike price, $\log K$, in order to achieve a higher accuracy of pricing European Vanilla Option. Our scheme of distributing the interpolation point is shown in Figure \ref{Chptr4:fig:ems_around_strike}. The idea can be explained as follows: We set the range of $[x_{\min},\, x_{\max}]$ and then use EMS to create $N$ interpolation points. We distribute the first $N/2$ points uniformly in $[x_{\rm min},\, \log(K)]$ and then the rest in $[\log(K),\, x_{\rm max}].$

In option trading the region of most interest is when the mean of the
stock prices is close to the strike price. Typically, the probability for a stock
to default or to be very far from the strike price is small. Therefore we define the region of interest as follows:
\begin{eqnarray}
\hat{x}_i\in[\hat{x}_{\min},\,\hat{x}_{\max}]:=[\,\log(K/20),\,\log(2K)\,].
\end{eqnarray}

Based on this region, we can measure the accuracy of our RBF-approximation. We use a set of evaluation points $\hat{x }_i^{\Delta x}$,
for which we will simply take the grid points
\begin{equation}
\hat{x}_i:=\hat{x}_i^{\Delta x}=\hat{x}_{\rm{min}}+j\Delta \hat{x},
\,j=0,1,2,\ldots,N_{\rm eval}-1. \label{Chptr4:eqn:evulation_pt_chosen}
\end{equation}
Here $\Delta\hat{x}= (\hat{x}_{\max}-\hat{x}_{\max})/(N_{\rm eval}-1)$ with
$x_{\min}\leq \hat{x}_{\min}\leq \hat{x}_{\max}\leq x_{\max}$ and
$N_{\rm eval}$ is the number of the evaluation points chosen.

It is also of great interest to measure the rate of convergence of our Cubic Spline-approximation scheme. By defining $\Delta t=1/M_0, $ where $M_0$ is the number of time steps and $\Delta x=1/N,$ where $N$ is the number of interpolation points, we assume that
\begin{eqnarray}\label{Chptr4:eqn:rateofConvergence_Max1}
E_\infty(\hat{x}_i,T)&=&C_t(\Delta t)^{R_t}+C_{x}(\Delta x)^{R_\infty}
\end{eqnarray}
for the max error and
\begin{eqnarray}\label{Chptr4:eqn:rateofConvergence_RMS1}
E_2(\hat{x}_i,T)&=&C_t(\Delta t)^{R_t}+C_{x}(\Delta x)^{R_2}
\end{eqnarray}
for the root-mean-square (rms) error. Here $E_\infty(\hat{x}_i,T)$ is the max error, $E_2(\hat{x}_i,T)$  is the rms error, $\hat{x}_i$ is $i^{th}$ evaluation point, $T$ is the maturity time, both $C_t$ and $C_{\hat{x}}$ are constants, $R_t$ is the rates of convergence in time and $R_\infty$ and $R_2$ are the rates of convergence in space. The formulae of calculating the max error and the rms errors are:
\begin{align}
E_\infty=\max_{0\leq i\leq N_{\rm eval}} |V(e^{\hat{x}_i},\tau)-\widetilde{u}(\hat{x}_i,\tau)|,
\label{linfty_error}
\end{align}
and
\begin{align}
E_2=\sqrt{\frac{1}{N_{\rm eval}}\sum_{0\leq i\leq
N_{\rm eval}}|V(e^{\hat{x}_i},\tau)-\widetilde{u}(\hat{x}_i,\tau)|^2} \label{Chptr4:eqn:rms_error}
\end{align}
respectively. Here $V(e^{\hat{x}},\tau)$ and $\widetilde{u}({\hat{x}},\tau)$ are the exact value and
approximate value at the point $({\hat{x}}, \tau)$ respectively.

Since we compare the accuracy of our Cubic Spline-approximation scheme with that of FDM and FEM, we use the relative error,
\begin{align}
E_{\rm
rel.}(\hat{x},\tau)=\frac{|V(e^{\hat{x}},\tau)-\widetilde{u}({\hat{x}},\tau)|}{V(e^{\hat{x}},\tau)},\label{Chptr4:eqn:rel_error}
\end{align}
as the measure of the accuracy.

It is known \citep{Merton:1976} that the analytical price of a European
call/put option in the Merton Jump-diffusion model is given by
\begin{eqnarray}\label{Chptr4:eqn:anlyMertonformula}
    &\ &V_{\rm MJ}(S_t,\tau,K,r,q,\sigma)\nonumber\\
    &\ &\nonumber\\
    &=&\sum_{k=0}^\infty\frac{e^{-\lambda(1+\eta)\tau}((\lambda(1+\eta)\tau)^k}{k!}V_{BS}(S_t,\tau,K,r_k,\sigma_k,q)
\end{eqnarray}
where $\tau=T-t$ is the time to maturity,
    $\eta=e^{\mu_J+\frac{\sigma_J^2}{2}}-1$ represents the expected
percentage change in the stock price originating from a jump,
$\sigma^2_k=\sigma^2+\frac{k\sigma_J^2}{T-t}$ the observed
volatility, $r_k=r-\lambda\eta+k\log(1+\eta)/(T-t) $, $q$ is the dividend and $V_{BS}$ the
Black-Scholes price of a call and put, computed as
\begin{eqnarray}
    &\ &V_{BS}(S_t,\tau,K,r_k,\sigma_k,q)\nonumber\\
    &=&\left\{\begin{array}{ll}
    S_te^{-q\tau}\Phi(d_{+,k})-Ke^{-r_k\tau}\Phi(d_{-,k})&\mbox{call option,}\\
    Ke^{-r_k\tau}\Phi(-d_{-,k})-S_te^{-q\tau}\Phi(-d_{+,k})&\mbox{put option,}\end{array}\right.\nonumber
\end{eqnarray}
where $\Phi(\cdot)$ is the cumulative normal distribution and
\begin{eqnarray}
    d_{+,k}&=\frac{\log(S_t/K)+(r_k-q+\sigma_k^2/2)\tau}{\sigma_k\sqrt{\tau}},
    \quad d_{-,k}=d_{+,k}-\sigma_k\sqrt{\tau}.\nonumber
\end{eqnarray}
For the derivation of $V_{\rm MJ}(S_t,\tau,K,r,q,\sigma)$, we shall refer to the reader to \citep{Merton:1976,Con_Tan:2004}.

In general, for models where the characteristic function of the L\'evy process is known, an analytical
solution of PIDE (\ref{Chptr4:eqn:PIDE_MJ2}) may be found using Fourier analysis \citep{Carr_Madan:1999, Lew:2001}. For the sake of simplicity and accuracy we propose Jackson \textsl{et al.}'s Fourier Space Time-Stepping method rather than Carr-Madan's Fast Fourier Transform (FFT) method \citep{Carr_Madan:1999} and Lewis's FFT method \citep{Lew:2001}. In brief, the idea of this method is based on the Fourier transform of the PIDE. By making use of FFT and inverse Fast Fourier transform ($\rm FFT^{-1}$), European Option price can be determined. The pricing formula of evaluating European option can be expressed as follows:
\begin{eqnarray}\label{Chptr4:eqn:FSTS_method}
 V_{\rm Kou}(S,\tau,K,r,\sigma,q)={\rm FFT^{-1}}[\,{\rm FFT} \,[V_{\rm Kou}(S,T)\,] e^{\psi \tau}\,],
\end{eqnarray}
where $\psi(z)$ is the characteristic function of the Kou model which can be defined as:
$$-\frac{\sigma^2z^2}{2}+iz\gamma_c+\lambda\big(\frac{p\alpha_1}{\alpha_1-iz}+\frac{(1-p)\alpha_2}{\alpha_2+iz}-1\big),$$
and $V_{\rm Kou}(S,T)$ is the payoff function (\ref{Chptr4:eqn:payoffCallPut}). For more details of this method, we shall refer the reader to \citep{Jac_Jai_Sur:2008}. This method has been reported
to have second order convergence in space in European cases.

Our RBF-algorithm for numerically solving (\ref{Chptr4:eqn:PIDE_MJ2}) with initial
condition (\ref{Chptr4:eqn:payoffCallPut}) runs as follows:
\begin{enumerate}
\item Find the RBF-approximation to the initial value $u(x,0)$  using ESM (see \ref{Chptr4:eqn:ems_pt_chosen}). This will provide us with a set of interpolation points $x_1,\ldots,x_n$, together with an initial vector $\pmb{\rho}(0)=\big(\rho_1(0),\ldots,\rho_{N}(0)\big)$.
\item Then use $\pmb{\rho}(0)$ as initial value for the system (\ref{Chptr4:eqn:simplifed_ODE1}). By using any stiff ODE solver,
we find out the $\pmb{\rho}(T)$ at time $T$.
\item Finally, substitute $\pmb{\rho}(T)$ back into $\sum_{j=1}^{N}\rho_j(T)\phi(|x-x_j|)$ to get
an approximate value of $u(x,T)$.
\end{enumerate}

In our numerical experiment we implement the algorithm in MATLAB
R2007b. We select our maximum and minimum logarithm price $x_{\min}$
$\big(\log(S_{\min})\big)$ and $x_{\max}$ $\big(\log(S_{\max})\big), $
as before, equal to $-10$ and $10$ respectively. Because of achieving more accurate approximation of the integral in (\ref{Chptr4:eqn:ODE_In_trunc}), we also set $\epsilon$ in both \ref{Chptr4:eqn:finite_rng_MJ} and \ref{Chptr4:eqn:finite_rng_Kou} to be $3.72\times10^{-40}$ for finding a finite computational interval $[y_{-\epsilon}, y_{\epsilon}]$. Moreover, we use function $quadgk$ which implements adaptive Gauss-Kronrod quadrature for computing equation (\ref{Chptr4:eqn:AdpLQ}) as well as function
$ode15s$ which implements backward differentiation formulas (BDFs) of order two for the calculation of equation
(\ref{Chptr4:eqn:simplifed_ODE1}). The main reason of choosing it is the following: According to \citep{Ise:2009} BDFs of orders 1 and 2 are A-stable (the stability region includes the entire left half complex plane). Since (\ref{Chptr4:eqn:simplifed_ODE1}) is stiff, according to Theorem 4.11 (The Dahlquist second barrier) of \citep{Ise:2009}, the highest order of an A stable multistep method\footnote{Multistep methods are used for the numerical solution of ordinary differential equations. Conceptually, a numerical method starts from an initial point and then takes a short step forward in time to find the next solution point. The process continues with subsequent steps to map out the solution.}, such as BDFs, is only two. We therefore conclude that our solution is second order convergence in time. By setting $R_t=2$, (\ref{Chptr4:eqn:rateofConvergence_Max1}) and (\ref{Chptr4:eqn:rateofConvergence_RMS1}) become
\begin{eqnarray}\label{Chptr4:eqn:rateofConvergence_Max_RMS2}
E_\infty(\hat{x}_i,T)&=&C_t(\Delta t)^{2}+C_{x}(\Delta x)^{R_\infty}\\
E_2(\hat{x}_i,T)&=&C_t(\Delta t)^{2}+C_{x}(\Delta x)^{R_2}
\end{eqnarray}
for the European option. This conclusion is in line with the finding of \citep{Pet_Lar_Mar_Per:2008}. In \citep{Pet_Lar_Mar_Per:2008} Pettersson \textsl{et al.} show that second order in time can be achieved in a European case due to the second order time-stepping scheme, BDFs of order 2. Although they solve Black Schole PDE rather than PIDE in their paper, an similar approach of solving European option like our approximation scheme is applied. The rest of this section, we numerically show that $R_\infty$ and $R_2$ are equal to 2. Besides this, we will numerically approximate $\Delta$ and $\Gamma$ and launch a comparison between our approximation scheme and FDM and FEM.

\begin{figure}[htbp]
\centering
\epsfig{file=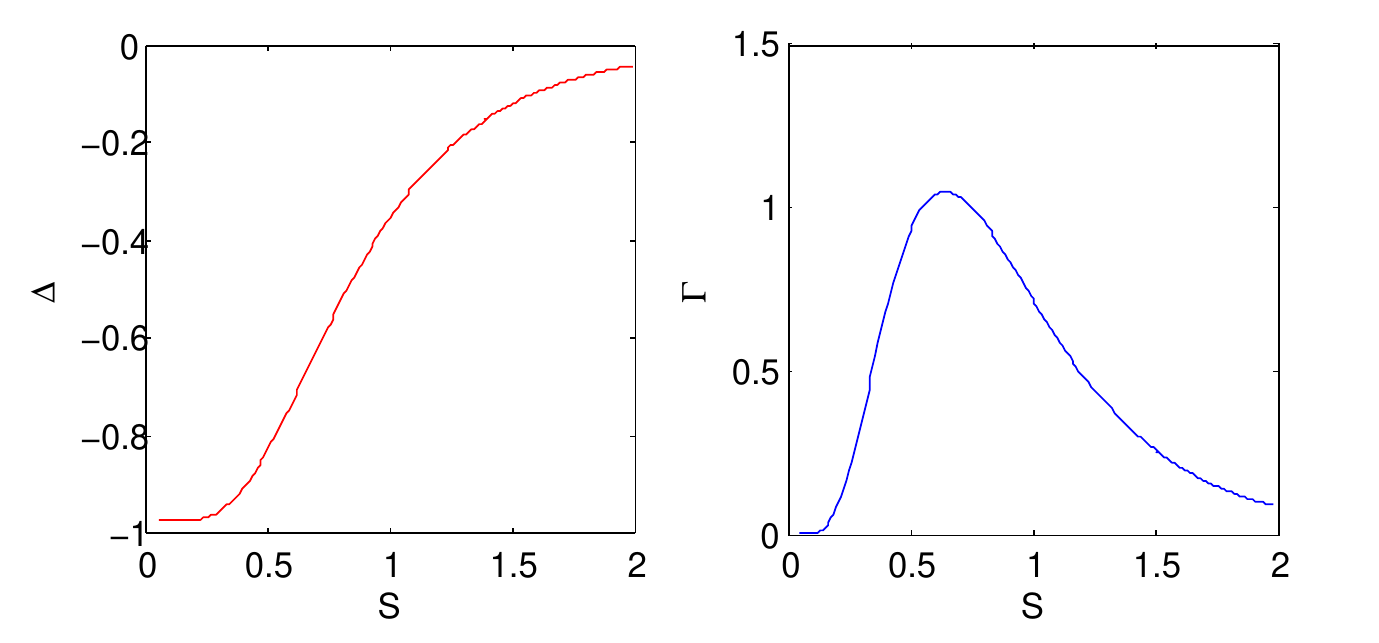}
\rule{30em}{0.5pt}
\caption{Put option Delta $\Delta$ (Left) and Gamma $\Gamma$ (Right) in the Black-Scholes Model. The number of the interpolation points is 3600. The number of evaluation points of a range of $S$ from 0.05 to 2 is 1950.The input parameters are provided in the caption to Table \ref{Chptr4:table:Euro_cubicBS3}.
}\label{Chptr4:fig:BSGreeks_Put}
\end{figure}

\begin{figure}[htbp]
\centering
\epsfig{file=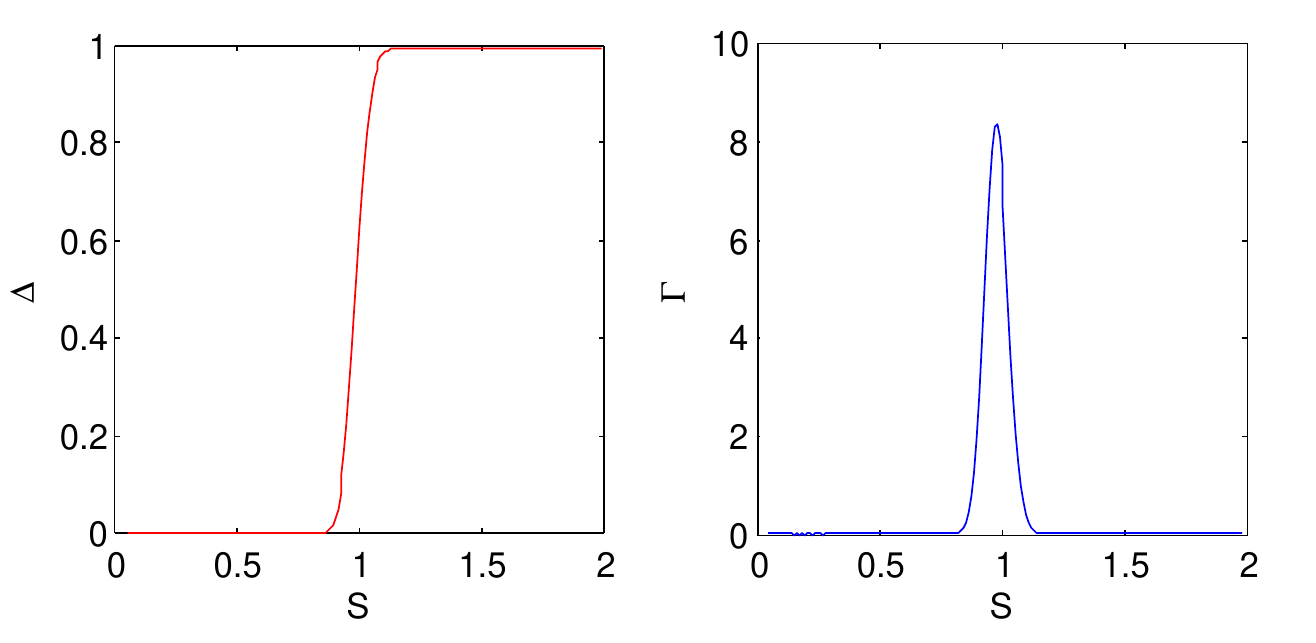}
\rule{30em}{0.5pt}
\caption{Call Option Delta $\Delta$ (Left) and Gamma $\Gamma$ (Right)in the Merton Jump-diffusion Model. The number of the interpolation points is 3600. The number of evaluation points of a range of $S$ from 0.05 to 2 is 1950.The input parameters are provided in the caption to Table \ref{Chptr4:table:Euro_cubicMJ2}.
}\label{Chptr4:fig:MertonGreeks_Call}
\end{figure}

\begin{figure}[htbp]
\centering
\epsfig{file=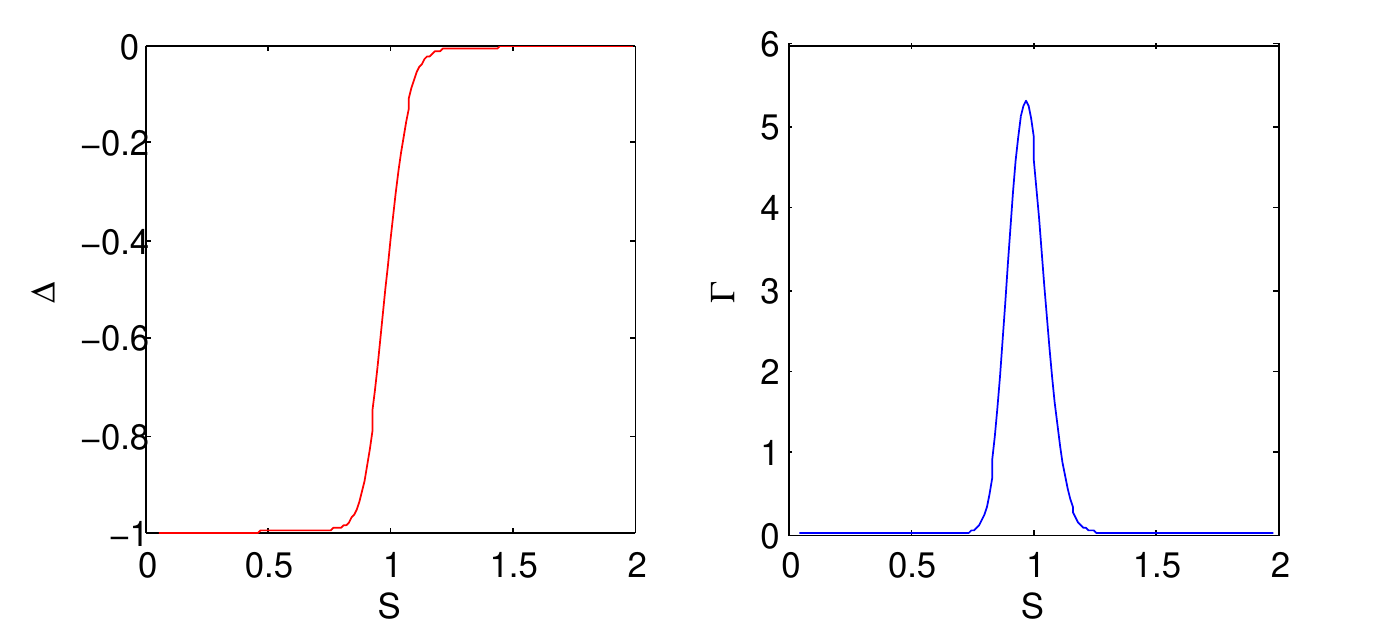}
\rule{30em}{0.5pt}
\caption{Put Option Delta $\Delta$ (Left) and Gamma $\Gamma$ (Right)in the Kou Jump-diffusion Model. The number of the interpolation points is 3600. The number of evaluation points of a range of $S$ from 0.05 to 2 is 1950.The input parameters are provided in the caption to Table \ref{Chptr4:table:Euro_cubicKou2}.
}\label{Chptr4:fig:KouGreeks_Put}
\end{figure}

All the parameters of all the tables except Table \ref{Chptr4:table:Euro_cubicBS3}, \ref{Chptr4:table:Euro_cubicMJ3} and \ref{Chptr4:table:Euro_cubicKou3} are chosen from different literature. The parameter $\sigma=1$ in Table \ref{Chptr4:table:Euro_cubicBS3}, \ref{Chptr4:table:Euro_cubicMJ3} and \ref{Chptr4:table:Euro_cubicKou3} is selected to stress our numerical algorithm. From Table \ref{Chptr4:table:Euro_cubicBS1} to \ref{Chptr4:table:Euro_cubicKou3}, $E_\infty$ and $E_2$ falls down when the number of the interpolation points $N$ increases. Our Cubic Spline approximation scheme can get second order convergence in space. This is due to the limited smoothness of the Cubic Spline which has second order of convergence (cf. \citep{Wendland:2005}). In Figure \ref{Chptr4:fig:BSGreeks_Put}, \ref{Chptr4:fig:MertonGreeks_Call} and \ref{Chptr4:fig:KouGreeks_Put}, oscillations do not occur around the strike $K$ for small $T$ when we approximate $\Delta$ and $\Gamma.$ In Table \ref{Chptr4:table:Euro_compareMJ1}, we compare the results of the FD used in Briani \textsl{et al.}'s paper \citep{Bri_Nat_Rus:2007} with those using our Cubic Spline approximation scheme. Our numerical approximation scheme can achieve lower $E_{\rm rel.}(\log S, T)$ than ARS-233 scheme and Explicit scheme. Table \ref{Chptr4:table:Euro_compareMJ1} and \ref{Chptr4:table:Euro_compareKou1} are other comparisons of the accuracy between our Cubic Spline approximation scheme and Almendral and Oosterlee's FD and FE with BDF2. To illustrate a fair comparison, we set our maximum and minimum logarithm price $x_{\min}$ and $x_{\max}$ same as
Almendral and Oosterlee proposed in their numerical experiments. Hence we set $[x_{\min}$ $x_{\max}]$ equal to [-4 4] and [-6 6] in the Merton model (Table \ref{Chptr4:table:Euro_compareMJ2}) and the Kou model (Table \ref{Chptr4:table:Euro_compareKou1}) respectively. Our Cubic Spline approximation scheme can attain lower $E_{\rm rel.}(\log S, T)$ than FD and FE with BDF2 in both the Merton and Kou cases.

\begin{table}[htbp]
\centering 
\begin{tabular}{c c c c c} 
\hline
\hline
N & $E_\infty(\hat{x}_i,T)$ & $R_\infty$& $E_2(\hat{x}_i,T)$ & $R_2$ \\ [0.5ex] 
\hline
\hline
100&	4.207101E-03&	N/A&	1.864736E-03&	N/A\\[0.5ex]
600&	1.195088E-04&	1.988&	5.143665E-05&	2.004\\[0.5ex]
1100&	3.554622E-05&	2.000&	1.528321E-05&	2.002\\[0.5ex]
1600&	1.679290E-05&	2.001&	7.219811E-06&	2.001\\[0.5ex]
2100&	9.745141E-06&	2.001&	4.189909E-06&	2.001\\[0.5ex]
2600&	6.354765E-06&	2.002&	2.732818E-06&	2.001\\[0.5ex]
3100&	4.468110E-06&	2.003&	1.921950E-06&	2.001\\[0.5ex]
3600&	3.311319E-06&	2.004&	1.424931E-06&	2.001\\[0.5ex]
\hline 
\hline
\end{tabular}
\medskip
\caption{$E_\infty$ and $E_2$ of the Cubic Spline
approximation for pricing of a European put under the Black-Scholes model are presented. $N$ is the number of the interpolation points. $\hat{x_i}=\log S_i $ is any evaluation points of a range of $S$ from 0.05 to 2 and the total numbers are 1950. T is the Time-to-maturity.
The parameters are: $r = 0.04,$ $q = 0,$ $\sigma = 0.29,$ $K=1$ and $T = 1.$
The parameters are taken from \cite{Gil_Car:2006}. The order of convergence is 2 in space. 	
} 
\label{Chptr4:table:Euro_cubicBS1}
\end{table}

\begin{table}[htbp]
\centering 
\begin{tabular}{c c c c c} 
\hline
\hline
N & $E_\infty(\hat{x}_i,T)$ & $R_\infty$& $E_2(\hat{x}_i,T)$ & $R_2$ \\ [0.5ex] 
\hline
\hline
100&	1.924131E-02&	N/A&	4.690135E-03&	N/A\\ [0.5ex]
600&	7.143939E-04&	1.838&	1.296858E-04&	2.003\\ [0.5ex]
1100&	2.171519E-04&	1.965&	3.870772E-05&	1.995\\ [0.5ex]
1600&	1.031950E-04&	1.986&	1.830673E-05&	1.998\\ [0.5ex]
2100&	6.002721E-05&	1.992&	1.063352E-05&	1.998\\ [0.5ex]
2600&	3.919766E-05&	1.995&	6.934013E-06&	2.002\\ [0.5ex]
3100&	2.758717E-05&	1.997&	4.877540E-06&	2.000\\ [0.5ex]
3600&	2.046213E-05&	1.998&	3.616699E-06&	2.000\\ [0.5ex]
\hline 
\hline
\end{tabular}
\medskip
\caption{$E_\infty$ and $E_2$ of the Cubic Spline
approximation for pricing of a European call under the Black-Scholes model are presented. $N$ is the number of the interpolation points. $\hat{x_i}=\log S_i $ is any evaluation points of a range of $S$ from 0.05 to 2 and the total numbers are 1950. T is the Time-to-maturity.
The parameters are: $r = 0.05,$ $q = 0,$ $\sigma = 0.2,$ $K=1$ and $T = 2.$
The parameters are taken from \cite{Shaw:1998}. The order of convergence is 2 in space. 	
} 
\label{Chptr4:table:Euro_cubicBS2}
\end{table}

\begin{table}[htbp]
\centering 
\begin{tabular}{c c c c c} 
\hline
\hline
N & $E_\infty(\hat{x}_i,T)$ & $R_\infty$& $E_2(\hat{x}_i,T)$ & $R_2$ \\ [0.5ex] 
\hline
\hline
100&	2.325676E-03&	0.000&	1.404611E-03&   N/A\\ [0.5ex]
600&	6.473617E-05&	1.999&	3.856043E-05&	2.007\\ [0.5ex]
1100&	1.923322E-05&	2.002&	1.145625E-05&	2.002\\ [0.5ex]
1600&	9.079037E-06&	2.003&	5.411921E-06&	2.001\\ [0.5ex]
2100&	5.265272E-06&	2.004&	3.140776E-06&	2.001\\ [0.5ex]
2600&	3.430306E-06&	2.006&	2.048580E-06&	2.001\\ [0.5ex]
3100&	2.406208E-06&	2.016&	1.441039E-06&	2.000\\ [0.5ex]
3600&	1.782442E-06&	2.007&	1.068202E-06&   2.002\\ [0.5ex]
\hline 
\hline
\end{tabular}
\medskip
\caption{$E_\infty$ and $E_2$ of the Cubic Spline
approximation for pricing of a European call under the Black-Scholes model are presented. $N$ is the number of the interpolation points. $\hat{x_i}=\log S_i $ is any evaluation points of a range of $S$ from 0.05 to 2 and the total numbers are 1950. T is the Time-to-maturity.
The parameters are: $r = 0.3,$ $q = 0.1,$ $\sigma = 1,$  $K=1$ and $T = 0.25$, whereas the parameter $\sigma=1$ is selected to stress our numerical algorithm.
 The order of convergence is 2 in space. 	
} 
\label{Chptr4:table:Euro_cubicBS3}
\end{table}

\begin{table}[htbp]
\centering 
\begin{tabular}{c c c c c} 
\hline
\hline
N & $E_\infty(\hat{x}_i,T)$ & $R_\infty$& $E_2(\hat{x}_i,T)$ & $R_2$ \\ [0.5ex] 
\hline
\hline
100&	1.428497E-02&	N/A&	3.749983E-03&	N/A\\ [0.5ex]
600&	4.642130E-04&	1.912&	1.011341E-04&	2.016\\ [0.5ex]
1100&	1.402519E-04&	1.975&	3.011378E-05&	1.999\\ [0.5ex]
1600&	6.640377E-05&	1.995&	1.423346E-05&	2.000\\ [0.5ex]
2100&	3.860331E-05&	1.995&	8.262241E-06&	2.000\\ [0.5ex]
2600&	2.518672E-05&	1.999&	5.389115E-06&	2.001\\ [0.5ex]
3100&	1.772559E-05&	1.997&	3.790660E-06&	2.000\\ [0.5ex]
3600&	1.314288E-05&	2.000&	2.810697E-06&	2.000\\ [0.5ex]
\hline 
\hline
\end{tabular}
\medskip
\caption{$E_\infty$ and $E_2$ of the Cubic Spline
approximation for pricing of a European call under the Merton Jump-diffusion model are presented. $N$ is the number of the interpolation points. $\hat{x_i}=\log S_i $ is any evaluation points of a range of $S$ from 0.05 to 2 and the total numbers are 1950. T is the Time-to-maturity.
The parameters are: $r = 0.05,$ $q = 0,$ $\sigma = 0.15,$ $\sigma_J = 0.45,$ $\mu_J = -0.9,$ $\lambda=0.1,$ $K=1$ and $T = 0.25.$
The parameters are taken from \citep{Ander_Andre:2000}. The order of convergence is 2 in space. 	
} 
\label{Chptr4:table:Euro_cubicMJ1}
\end{table}

\begin{table}[htbp]
\centering 
\begin{tabular}{c c c c c} 
\hline
\hline
N & $E_\infty(\hat{x}_i,T)$ & $R_\infty$& $E_2(\hat{x}_i,T)$ & $R_2$ \\ [0.5ex] 
\hline
\hline
100&	1.956920E-02&	N/A&	4.723349E-03&	N/A\\ [0.5ex]
600&	7.326011E-04&	1.833&	1.305576E-04&	2.003\\ [0.5ex]
1100&	2.240092E-04&	1.955&	3.898655E-05&	1.994\\ [0.5ex]
1600&	1.069094E-04&	1.974&	1.844062E-05&	1.998\\ [0.5ex]
2100&	6.223777E-05&	1.990&	1.071235E-05&	1.997\\ [0.5ex]
2600&	4.062560E-05&	1.997&	6.985440E-06&	2.002\\ [0.5ex]
3100&	2.859186E-05&	1.997&	4.913762E-06&	2.000\\ [0.5ex]
3600&	2.121748E-05&	1.995&	3.643595E-06&	2.000\\ [0.5ex]
\hline 
\hline
\end{tabular}
\medskip
\caption{$E_\infty$ and $E_2$ of the Cubic Spline
approximation for pricing of a European put under the Merton Jump-diffusion model are presented. $N$ is the number of the interpolation points. $\hat{x_i}=\log S_i $ is any evaluation points of a range of $S$ from 0.05 to 2 and the total numbers are 1950. T is the Time-to-maturity.
The parameters are: $r = 0.05,$ $q = 0.02,$ $\sigma = 0.15,$ $\sigma_J = 0.4,$ $\mu_J = -1.08,$ $\lambda = 0.1,$ $K=1$ and $T = 0.1.$
The parameters are taken from \citep{Ander_Andre:2000}. The order of convergence is 2 in space. 	
} 
\label{Chptr4:table:Euro_cubicMJ2}
\end{table}

\begin{table}[htbp]
\centering 
\begin{tabular}{c c c c c} 
\hline
\hline
N & $E_\infty(\hat{x}_i,T)$ & $R_\infty$& $E_2(\hat{x}_i,T)$ & $R_2$ \\ [0.5ex] 
\hline
\hline
100&	1.026524E-03&	N/A&	7.090253E-04&	N/A\\ [0.5ex]
600&	2.819557E-05&	2.006&	1.945356E-05&	2.007\\ [0.5ex]
1100&	8.415823E-06&	1.995&	5.762520E-06&	2.007\\ [0.5ex]
1600&	3.999351E-06&	1.986&	2.712396E-06&	2.011\\ [0.5ex]
2100&	2.373272E-06&	1.919&	1.559774E-06&	2.035\\ [0.5ex]
2600&	1.601472E-06&	1.842&	1.004746E-06&	2.059\\ [0.5ex]
3100&	1.136188E-06&	1.951&	7.021072E-07&	2.038\\ [0.5ex]
3600&	8.358248E-07&	2.053&	5.221973E-07&	1.980\\ [0.5ex]
\hline
\hline
\end{tabular}
\medskip
\caption{$E_\infty$ and $E_2$ of the Cubic Spline
approximation for pricing of a European call under the Merton Jump-diffusion model are presented. $N$ is the number of the interpolation points. $\hat{x_i}=\log S_i $ is any evaluation points of a range of $S$ from 0.05 to 2 and the total numbers are 1950. T is the Time-to-maturity.
The parameters are: $r = 0.05,$ $q = 0.01,$ $\sigma = 1,$ $\sigma_J = 0.6,$ $\mu_J = -1.08,$ $\lambda = 0.1,$ $K=1$ and $T = 1$, whereas the parameter $\sigma=1$ is selected to stress our numerical algorithm.
The order of convergence is 2 in space. 	
} 
\label{Chptr4:table:Euro_cubicMJ3}
\end{table}

\begin{table}[htbp]
\centering 
\begin{tabular}{c c c c c} 
\hline
\hline
N & $E_\infty(\hat{x}_i,T)$ & $R_\infty$& $E_2(\hat{x}_i,T)$ & $R_2$ \\ [0.5ex] 
\hline
\hline
100&	1.239165E-02&	N/A&	3.422908E-03&	N/A\\ [0.5ex]
600&	3.932126E-04&	1.926&	9.440247E-05&	2.004\\ [0.5ex]
1100&	1.179555E-04&	1.986&	2.808850E-05&	2.000\\ [0.5ex]
1600&	5.589111E-05&	1.993&	1.327392E-05&	2.000\\ [0.5ex]
2100&	3.246588E-05&	1.998&	7.705266E-06&	2.000\\ [0.5ex]
2600&	2.118103E-05&	2.000&	5.025765E-06&	2.001\\ [0.5ex]
3100&	1.490021E-05&	2.000&	3.535171E-06&	2.000\\ [0.5ex]
3600&	1.105067E-05&	1.999&	2.621377E-06&	2.000\\ [0.5ex]
\hline 
\hline
\end{tabular}
\medskip
\caption{$E_\infty$ and $E_2$ of the Cubic Spline
approximation for pricing of a European put under the Kou Jump-diffusion model are presented. $N$ is the number of the interpolation points. $\hat{x_i}=\log S_i $ is any evaluation points of a range of $S$ from 0.05 to 2 and the total numbers are 1950. T is the Time-to-maturity.
The parameters are: $r = 0,$ $q = 0,$ $\sigma = 0.2,$ $\alpha_1=3,$ $\alpha_2 = 2,$ $\lambda = 0.2,$ $p = 0.5,$ $K=1$ and $T = 0.2.$
The parameters are taken from \citep{Alm_Oos:2005}. The order of convergence is 2 in space. 	
} 
\label{Chptr4:table:Euro_cubicKou1}
\end{table}

\begin{table}[htbp]
\centering 
\begin{tabular}{c c c c c} 
\hline
\hline
N & $E_\infty(\hat{x}_i,T)$ & $R_\infty$& $E_2(\hat{x}_i,T)$ & $R_2$ \\ [0.5ex] 
\hline
\hline
100&	1.433875E-02&	N/A&	3.766745E-03&	N/A\\ [0.5ex]
600&	4.665677E-04&	1.912&	1.022079E-04&	2.013\\ [0.5ex]
1100&	1.404381E-04&	1.981&	3.043034E-05&	1.999\\ [0.5ex]
1600&	6.660275E-05&	1.991&	1.438190E-05&	2.000\\ [0.5ex]
2100&	3.868283E-05&	1.998&	8.348098E-06&	2.000\\ [0.5ex]
2600&	2.522395E-05&	2.002&	5.444331E-06&	2.001\\ [0.5ex]
3100&	1.773247E-05&	2.003&	3.828943E-06&	2.001\\ [0.5ex]
3600&	1.314079E-05&	2.004&	2.838628E-06&	2.001\\ [0.5ex]
\hline 
\hline
\end{tabular}
\medskip
\caption{$E_\infty$ and $E_2$ of the Cubic Spline
approximation for pricing of a European call under the Kou Jump-diffusion model are presented. $N$ is the number of the interpolation points. $\hat{x_i}=\log S_i $ is any evaluation points of a range of $S$ from 0.05 to 2 and the total numbers are 1950. T is the Time-to-maturity.
The parameters are: $r = 0.05,$ $q = 0,$ $\sigma = 0.15,$ $\alpha_1=3.0465,$ $\alpha_2 = 3.0465,$ $\lambda = 0.1,$ $p = 0.3445,$ $K=1$ and $T = 0.25.$
The parameters are taken from \citep{Carr_Ani_JumpDiff:2007}. The order of convergence is 2 in space. 	
} 
\label{Chptr4:table:Euro_cubicKou2}
\end{table}

\begin{table}[htbp]
\centering 
\begin{tabular}{c c c c c} 
\hline
\hline
N & $E_\infty(\hat{x}_i,T)$ & $R_\infty$& $E_2(\hat{x}_i,T)$ & $R_2$ \\ [0.5ex] 
\hline
\hline
100&	1.080306E-03&	N/A&	7.074108E-04&	N/A\\ [0.5ex]
600&	2.973137E-05&	2.005&	1.940773E-05&	2.007\\ [0.5ex]
1100&	8.870629E-06&	1.995&	5.757611E-06&	2.005\\ [0.5ex]
1600&	4.229400E-06&	1.977&	2.712641E-06&	2.009\\ [0.5ex]
2100&	2.490583E-06&	1.947&	1.567674E-06&	2.016\\ [0.5ex]
2600&	1.674611E-06&	1.859&	1.014582E-06&	2.037\\ [0.5ex]
3100&	1.191565E-06&	1.935&	7.096338E-07&	2.032\\ [0.5ex]
3600&	9.018770E-07&	1.863&	5.232205E-07&	2.038\\ [0.5ex]
\hline 
\hline
\end{tabular}
\medskip
\caption{$E_\infty$ and $E_2$ of the Cubic Spline
approximation for pricing of a European put under the Kou Jump-diffusion model are presented. $N$ is the number of the interpolation points. $\hat{x_i}=\log S_i $ is any evaluation points of a range of $S$ from 0.05 to 2 and the total numbers are 1950. T is the Time-to-maturity.
The parameters are: $r = 0.04,$ $q = 0.03,$ $\sigma = 1,$ $\alpha_1=4,$ $\alpha_2 = 4,$ $\lambda = 0.3,$ $p = 0.6$ $K=1$ and $T = 1,$ whereas the parameter $\sigma=1$ is selected to stress our numerical algorithm.
The order of convergence is 2 in space. 	
} 
\label{Chptr4:table:Euro_cubicKou3}
\end{table}
\begin{table}[htbp]
\centering 
\begin{tabular}{c c c c c c} 

& $\,$  & \multicolumn{2}{c}{Explicit scheme} & \multicolumn{2}{c}{ARS-233 Scheme}\\
\hline
\hline
& N & $Value$&$E_{\rm rel.}(\log S,T)$&$Value$&$E_{\rm rel.}(\log S,T)$\\ [0.5ex] 
\hline
\hline
Call &1024&13.286915&5.175624E-03&13.287427&5.214358E-03\\[0.5ex]
Put  &1024&8.319940&2.57797E-03&8.326102&1.839249E-03\\[0.5ex]
\hline
&&&&\\
& $\,$  & \multicolumn{2}{c}{Cubic Spline} & \multicolumn{2}{c}{N/A}\\[0.5ex]
\hline
\hline
& N & $Value$&$E_{\rm rel.}(\log S,T)$&$Value$&$E_{\rm rel.}(\log S,T)$\\ [0.5ex] 
\hline
\hline
Call &1024&13.219358&6.489263E-05&N/A&N/A\\[0.5ex]
Put  &1024&8.342301&1.027679E-04&N/A&N/A\\[0.5ex]
\hline 
\hline
\end{tabular}
\medskip
\caption{Comparison between Explicit scheme (\cite{Bri_Nat_Rus:2007}),  ARS-233 Scheme (\cite{Bri_Nat_Rus:2007})and Cubic Spline interpolation scheme
in evaluating of European call/put under the Merton Jump-diffusion Model.
The input parameters are: $r = 0.05,$ $q = 0,$ $\sigma = 0.2,$ $\sigma_J = 0.8,$ $\mu_J = 0,$ $\lambda = 0.1,$ $K=100,$  $T = 1,$ and $x=\log 100$. Reference prices of 13.218501 (call) and 8.341444 (put) and parameters from \cite{Bri_Nat_Rus:2007}.
} 
\label{Chptr4:table:Euro_compareMJ1}
\end{table}

\begin{table}[htbp]
\centering 
\begin{tabular}{c c c c c } 

$\,$  & \multicolumn{2}{c}{FD with BDF2} & \multicolumn{2}{c}{FE with BDF2} \\
\hline
\hline
N & $Value$&$E_{\rm rel.}(\log S,T)$&$Value$&$E_{\rm rel.}(\log S,T)$\\ [0.5ex] 
1025&9.411968E-02&1.682457e-04&9.412972E-02&6.165536E-05\\[0.5ex]
\hline
&&&&\\
$\,$  &\multicolumn{2}{c}{Cubic Spline}& \multicolumn{2}{c}{N/A} \\[0.5ex]
\hline
\hline
N & $Value$&$E_{\rm rel.}(\log S,T)$&$Value$&$E_{\rm rel.}(\log S,T)$\\ [0.5ex]
1025&9.413023E-02&5.621522E-005& N/A&N/A\\[0.5ex]
\hline 
\hline
\end{tabular}
\medskip
\caption{Comparison of FD with BDF2 (\cite{Alm_Oos:2005}), FE with BDF2 (\cite{Alm_Oos:2005})
and Cubic Spline interpolation scheme in evaluating of a European call (put) under the Merton Jump-diffusion Model.
The input parameters are: $r = 0,$ $q = 0,$ $\sigma = 0.2,$ $\sigma_J = 0.5,$
$\mu_J = 0,$ $\lambda = 0.1,$ $K=1,$  $T = 1,$ and $S=1$.
Reference prices of 0.094135525 for both call and put and parameters from \cite{Alm_Oos:2005}.
} 
\label{Chptr4:table:Euro_compareMJ2}
\end{table}

\begin{table}[htbp]
\centering 
\begin{tabular}{c c c c c } 

$\,$  & \multicolumn{2}{c}{FD with BDF2} & \multicolumn{2}{c}{FE with BDF2} \\
\hline
\hline
N & $Value$&$E_{\rm rel.}(\log S,T)$&$Value$&$E_{\rm rel.}(\log S,T)$\\ [0.5ex] 
513&4.240E-02&6.346096E-03&4.24579E-02&5.1285862E-03\\[0.5ex]
\hline
&&&&\\
$\,$  &\multicolumn{2}{c}{Cubic Spline}& \multicolumn{2}{c}{N/A} \\[0.5ex]
\hline
\hline
N & $Value$&$E_{\rm rel.}(\log S,T)$&$Value$&$E_{\rm rel.}(\log S,T)$\\ [0.5ex]
513&4.254583E-02&3.061686E-03& N/A&N/A\\[0.5ex]
\hline 
\hline
\end{tabular}
\medskip
\caption{Comparison of FD with BDF2 (\cite{Alm_Oos:2005}), FE with BDF2 (\cite{Alm_Oos:2005})
and Cubic Spline interpolation scheme in evaluating of a European call (put) under the Kou Jump-diffusion Model.
The input parameters are: $r = 0,$ $q = 0,$ $\sigma = 0.2,$ $\alpha_1 = 3,$
$\alpha_2 = 2,$ $\lambda = 0.2,$ $p=0.5,$ $K=1,$  $T = 0.2,$ and $S=1$.
Reference prices of 0.0426761 for both call and put and parameters from \cite{Alm_Oos:2005}.
} 
\label{Chptr4:table:Euro_compareKou1}
\end{table}

\clearpage
\subsection{American Vanilla Put Options}\label{Chptr4:subsection:results:Amer_vanilla}
In this section we adapt an RBF-algorithm to compute American
put-option prices. We then compare the option prices obtained from
our RBF-algorithm with the Jackson \text{et al.} FST methods of \citep{Jac_Jai_Sur:2008}. As mentioned in Section
\ref{Chptr4:section:PIDEJumpD}, an American put option problem is a
free boundary problem because of the possibility of  early exercise
at any point during its life, leading to the free boundary
condition:
$$u(x,\tau)=\max\big(K-e^x, u(x,\tau)\big).$$
Together with the smooth pasting condition mentioned in section
\ref{Chptr4:section:PIDEJumpD}, this uniquely determines the exercise
boundary.

The Jackson \text{et al.} FST methods suggest that their solutions can achieve second order in space when they implement their methods to price American put options. They implement their methods in the context of the LCP. As we have seen in Section \ref{Chptr4:section:PIDEJumpD}, the value of an American option $u(\tau,x)$ is always greater than or equal to the payoff function $G(e^x)$. To numerically keep the condition $u(\tau,x)-G(e^x)\geq 0$ to be continuously held (see Section \ref{Chptr4:section:PIDEJumpD}), this can be achieved when boundary conditions are applied. The numerical algorithm for this idea can be defined as follows:
\begin{eqnarray}\label{Chptr4:eqn:FSTS_method_Amer}
 &\,&V(S,(m+1)\Delta t,K,r,\sigma,q)\nonumber\\
 &=&\max\{{\rm FFT^{-1}}[\,{\rm FFT} \,[V(S,m\Delta t,K,r,\sigma,q)\,] e^{\psi \Delta t}\,],G(e^x)\}
\end{eqnarray}
where time interval $\Delta t$ is obtained by dividing time-to-maturity $T$ by the total number $M$, $m\Delta$
is the time-step, where $m\in\{0,1,2,\ldots,M-1\}$, $\psi(z)$ is the characteristic function of the Merton/Kou models, $V(S,(m+1)\Delta t,K,r,\sigma,q)$ is the American put price at time $(m+1)\Delta t$ and the payoff condition $G(e^x)$ is equal to $\max(K-e^x,0).$ These methods also are required to swap between real and Fourier spaces at each time-step when the American option prices are calculated at each time interval. This is due to no convenient representation of the $\max(.,.)$ operator in Fourier space. For the full schematic and numerical description of this method, we refer readers to \citep{Jac_Jai_Sur:2008}.

As before, we use ESM to approximate
$u(x,0)=\max(K-e^x,0)$ and then continue to work with the
interpolation points found at $\tau=0$. The algorithm now reads as
follows:
\begin{enumerate}
\item Divide time-to-maturity $T$ by total numbers of time-steps $M$ to
obtain time interval $\Delta t$ and create a list of equally spaced
time-points $m\Delta t$, $m\in\{0,1,2,\ldots,M-1\}.$
\item Find the RBF-approximation to the initial value $u(x,0)$ using ESM. This will provide us with a set of
interpolation points $x_1,\ldots, x_n$, together with an initial
vector $\pmb{\rho}(0)=\big(\rho_1(0),\ldots,\rho_{N}(0)\big)$.
\item Assume we have already determined $\pmb{\rho}(m\Delta t)$ (if $m=0$, we have $\pmb{\rho}(0)$) in equation (\ref{Chptr4:eqn:simplifed_ODE1}).
Solve the system of (stiff) ODEs to find $\pmb{\rho}\big((m+1)\Delta
t\big)$ at the next successive time-step, $(m+1)\Delta t$.

\item Then at time $(m+1 )\Delta t$, for each interpolation point $x_i$, define
$$u\big(x_i,(m+1)\Delta
t\big)=\max\big((K-e^{x_i}),\,\sum_{j=1}^{N}\rho_j\big((m+1)\Delta
t\big)\phi(|x_i-x_j|)\big).$$

\item Find a new vector $\pmb \rho\big((m+1)\Delta t\big)$ such that $u\big(x_i,(m+1)\Delta
t\big)=\sum_{j=1}^{N}\rho_j\big((m+1)\Delta t\big)\phi(|x_i-x_j|)$
for all $i$.

\item Repeat Step 3.) to 5.) until $m=M-1$.
\item Finally, substitute $\pmb{\rho}(T)$ back into $\sum_{j=1}^{N}\rho_j(T)\phi(|x-x_j|)$ to get
an approximate value of $u(x,T)$.
\end{enumerate}

The settings of our numerical experiment are the same as those in
section \ref{Chptr4:subsection:results:Euro_vanilla}. We also calculate the rate of convergence in time. If we hold $\Delta x$ constant, (\ref{Chptr4:eqn:rateofConvergence_Max1}) and (\ref{Chptr4:eqn:rateofConvergence_RMS1})
become
\begin{eqnarray}\label{Chptr4:eqn:rateofConvergence_Max}
E_\infty(\hat{x}_i,T)&=&C_t(\Delta t)^{R_t}
\end{eqnarray}
and
\begin{eqnarray}\label{Chptr4:eqn:rateofConvergence_RMS}
E_2(\hat{x}_i,T)&=&C_t(\Delta t)^{R_t}
\end{eqnarray}
respectively.

\begin{figure}[htbp]
\centering
\epsfig{file=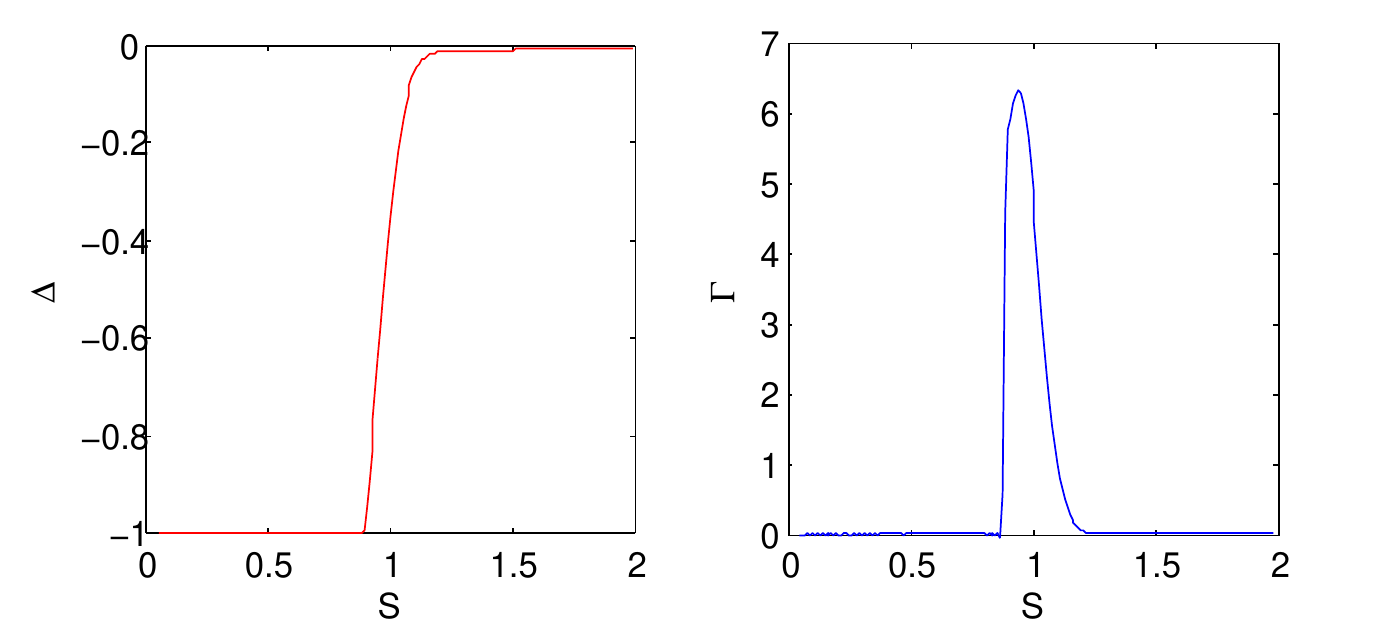}
\rule{30em}{0.5pt}
\caption{Put option Delta $\Delta$ (Left) and Gamma $\Gamma$ (Right) in the Merton Jump diffusion Model. The number of the interpolation points $N$ is 1800 and the number of time steps $M_0$ is $640$. The number of evaluation points of a range of $S$ from 0.05 to 2 is 1950.The input parameters are provided in the caption to Table \ref{Chptr4:table:Amer_cubicMerton1}.
}\label{Chptr4:fig:MertonGreeks_Put_Amer1}
\end{figure}

\begin{figure}[htbp]
\centering
\epsfig{file=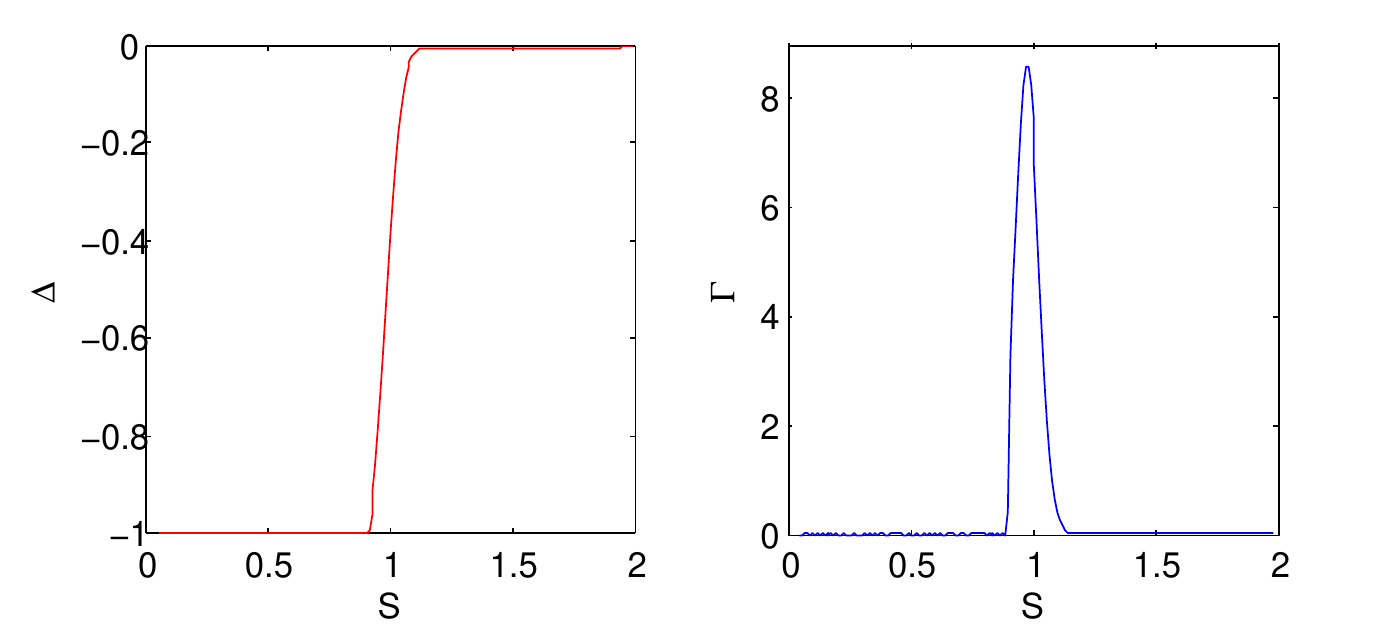}
\rule{30em}{0.5pt}
\caption{Put option Delta $\Delta$ (Left) and Gamma $\Gamma$ (Right) in the Merton Jump diffusion Model. The number of the interpolation points $N$ is 1800 and the number of time steps $M_0$ is $640$. The number of evaluation points of a range of $S$ from 0.05 to 2 is 1950.The input parameters are provided in the caption to Table \ref{Chptr4:table:Amer_cubicMerton2}.
}\label{Chptr4:fig:MertonGreeks_Put_Amer2}
\end{figure}

\begin{figure}[htbp]
\centering
\epsfig{file=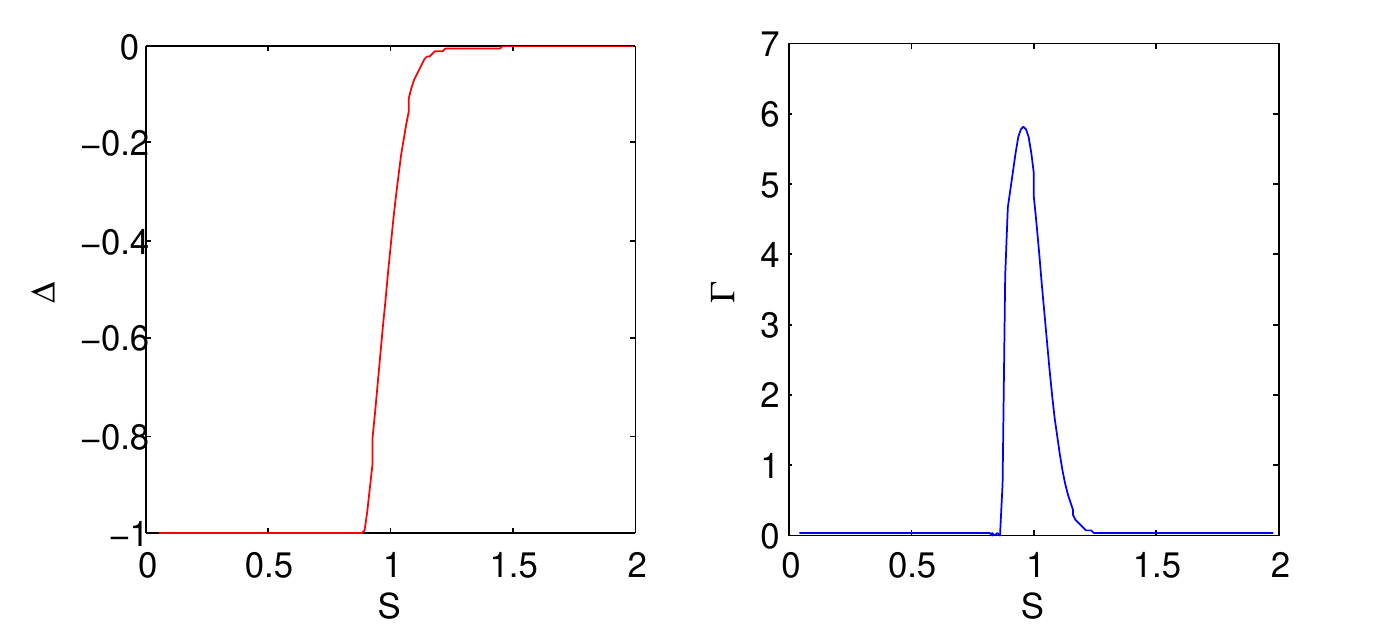}
\rule{30em}{0.5pt}
\caption{Put Option Delta $\Delta$ (Left) and Gamma $\Gamma$ (Right)in the Kou Jump-diffusion Model. The number of the interpolation points $N$ is 1800 and the number of time steps $M_0$ is $640$. The number of evaluation points of a range of $S$ from 0.05 to 2 is 1950.The input parameters are provided in the caption to Table \ref{Chptr4:table:Amer_cubicKou2}.
}\label{Chptr4:fig:KouGreeks_Put_Amer2}
\end{figure}

\begin{figure}[htbp]
\centering
\epsfig{file=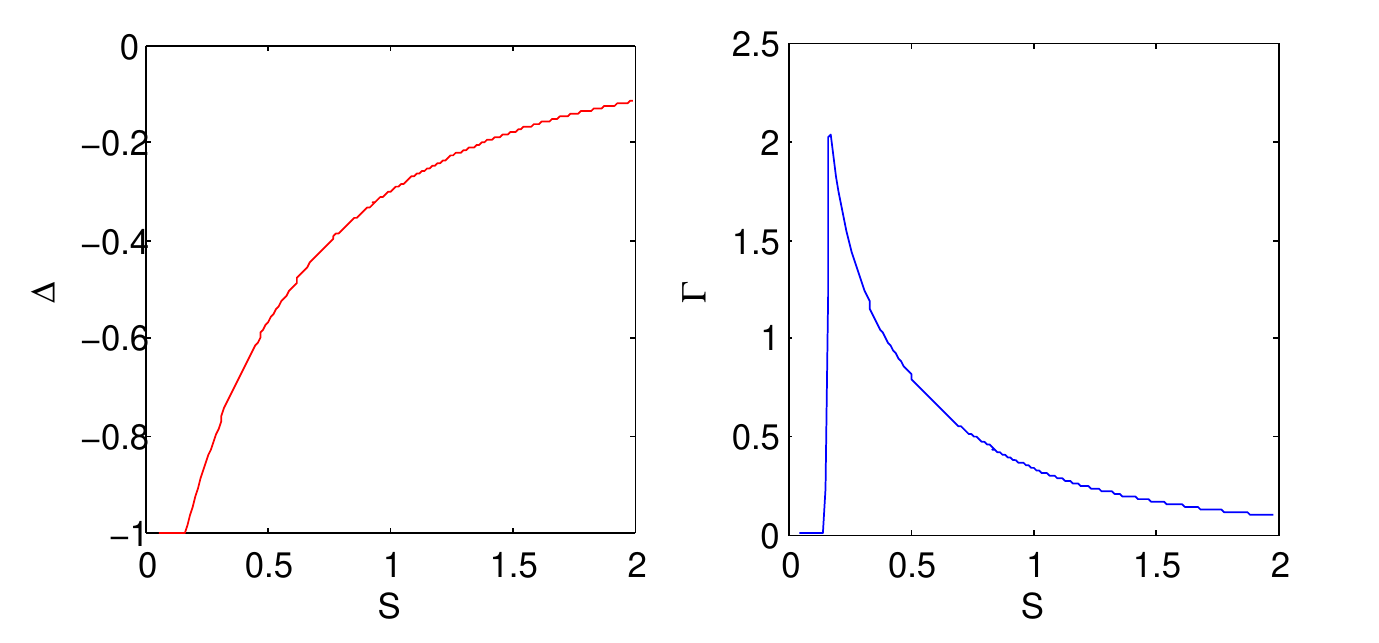}
\rule{30em}{0.5pt}
\caption{Put Option Delta $\Delta$ (Left) and Gamma $\Gamma$ (Right)in the Kou Jump-diffusion Model. The number of the interpolation points $N$ is 1800 and the number of time steps $M_0$ is $640$. The number of evaluation points of a range of $S$ from 0.05 to 2 is 1950.The input parameters are provided in the caption to Table \ref{Chptr4:table:Amer_cubicKou3}.
}\label{Chptr4:fig:KouGreeks_Put_Amer3}
\end{figure}

The results from Table \ref{Chptr4:table:Amer_cubicMerton1} to \ref{Chptr4:table:Amer_cubicKou3} suggest that our Cubic Spline approximation method for pricing of American put options is second order in spatial variables and first order in time variables when the number of interpolation numbers $N$ and the number of time-steps $M_0$ are twofold and fourfold respectively. In table \ref{Chptr4:table:Amer_cubicMJ_timstep} and \ref{Chptr4:table:Amer_cubicKou_timstep}, we implement our own BDF2 with fixed time steps rather than using $ode15s$ with variable time steps. From these two tables, we can achieve first order in time variables when the number of interpolation numbers $N$ is held constant and the number of time-steps $M_0$ is quadrupled. Moreover, from Figure \ref{Chptr4:fig:MertonGreeks_Put_Amer1} to \ref{Chptr4:fig:KouGreeks_Put_Amer3}, oscillations do not occur around the strike $K$ for small or big $T $ when we approximate $\Delta$ and $\Gamma.$


\begin{table}[htbp]
\centering 
\begin{tabular}{c c c c c c} 
\hline
\hline
N & $M_0$ & $E_\infty(\hat{x}_i,T)$ & $R_\infty$ & $E_2(\hat{x}_i,T)$ & $R_2$ \\ [0.5ex] 
\hline
\hline
225&	10&	   2.368536E-03&	N/A  &	1.007946E-03&	N/A  \\ [0.5ex]
450&	40&	   7.746936E-04&	1.612&	2.740154E-04&	1.879\\ [0.5ex]
900&	160&   2.260415E-04&	1.777&	6.969946E-05&	1.975\\ [0.5ex]
1800&	640&   6.362341E-05&	1.829&	1.888980E-05&	1.884\\ [0.5ex]
3600&	2560&  1.613907E-05&    1.979&	4.715908E-06&	2.002\\ [0.5ex]
\hline 
\hline
\end{tabular}
\medskip
\caption{$E_\infty$ and $E_2$ of the Cubic Spline approximation for pricing of an American put under the Merton model are presented. $N$ is the number of the interpolation points. $M_0$ is the number of the time steps. $\hat{x_i}=\log S_i $ is any evaluation points of a range of $S$ from 0.05 to 2 and the total numbers are 1950. T is the Time-to-maturity. The parameters are: $r = 0.05,$ $q = 0,$ $\sigma = 0.15,$ $\sigma_J = 0.45,$ $\mu_J = -0.9,$ $\lambda=0.1,$ $K=1$ and $T = 0.25.$
 The parameters are taken from \citep{Ander_Andre:2000}. The order of convergence is 2 in space and 1 in time. 	
} 
\label{Chptr4:table:Amer_cubicMerton1}
\end{table}

\begin{table}[htbp]
\centering 
\begin{tabular}{c c c c c c} 
\hline
\hline
N & $M_0$ & $E_\infty(\hat{x}_i,T)$ & $R_\infty$ & $E_\infty(\hat{x}_i,T)$ & $R_2$ \\ [0.5ex] 
\hline
\hline
225&	10&	3.401417E-03&	N/A&	7.995993E-04&	N/A\\ [0.5ex]
450&	40&	1.318325E-03&	1.367&	2.451148E-04&	1.706\\ [0.5ex]
900&	160&	3.744579E-04&	1.816&	6.873071E-05&	1.834\\ [0.5ex]
1800&	640&	1.055849E-04&	1.826&	1.927219E-05&	1.834\\ [0.5ex]
3600&	2560&	2.823205E-05&	1.903&	5.121082E-06 &	1.912\\ [0.5ex]
\hline 
\hline
\end{tabular}
\medskip
\caption{$E_\infty$ and $E_2$ of the Cubic Spline approximation for pricing of an American put under the Merton model are presented. $N$ is the number of the interpolation points. $M_0$ is the number of the time steps. $\hat{x_i}=\log S_i $ is any evaluation points of a range of $S$ from 0.05 to 2 and the total numbers are 1950. T is the Time-to-maturity.
The parameters are: $r = 0.05,$ $q = 0.02,$ $\sigma = 0.15,$ $\sigma_J = 0.4,$ $\mu_J = -1.08,$ $\lambda=0.1,$ $K=1$ and $T = 0.1.$
 The parameters are taken from \citep{Ander_Andre:2000}. The order of convergence is 2 in space and 1 in time. 	
} 
\label{Chptr4:table:Amer_cubicMerton2}
\end{table}
\begin{table}[htbp]
\centering 
\begin{tabular}{c c c c c c} 
\hline
\hline
N & $M_0$ & $E_\infty(\hat{x}_i,T)$ & $R_\infty$ & $E_2(\hat{x}_i,T)$ & $R_2$ \\ [0.5ex] 
\hline
\hline
225&	10&	4.935878E-03&	N/A  &	1.613323E-03&	N/A\\ [0.5ex]
450&	40&	1.236617E-03&	1.997&	3.725615E-04&	2.114\\ [0.5ex]
900&	160&	3.093198E-04&	1.999&	9.101657E-05&	2.033\\ [0.5ex]
1800&	640&	7.734030E-05&	2.000&	2.133679E-05&	2.093\\ [0.5ex]
3600&   2560&   1.932168E-005&  2.001&  5.074520E-06&   2.072\\ [0.5ex]
\hline 
\hline
\end{tabular}
\medskip
\caption{$E_\infty$ and $E_2$ of the Cubic Spline approximation for pricing of an American put under the Merton model are presented. $N$ is the number of the interpolation points. $M_0$ is the number of the time steps. $\hat{x_i}=\log S_i $ is any evaluation points of a range of $S$ from 0.05 to 2 and the total numbers are 1950. T is the Time-to-maturity.
The parameters are: $r = 0.05,$ $q = 0.01,$ $\sigma = 1,$ $\sigma_J = 0.6,$ $\mu_J = -1.08,$ $\lambda=0.1,$ $K=1$ and $T = 1,$ whereas the parameter $\sigma=1$ is selected to stress our numerical algorithm.
The order of convergence is 2 in space and 1 in time. 	
} 
\label{Chptr4:table:Amer_cubicMerton3}
\end{table}
\begin{table}[htbp]
\centering 
\begin{tabular}{c c c c c c} 
\hline
\hline
N & $M_0$ & $E_\infty(\hat{x}_i,T)$ & $R_\infty$ & $E_2(\hat{x}_i,T)$ & $R_2$ \\ [0.5ex] 
\hline
\hline
225&	10&	1.508321E-03&	N/A&	5.589125E-04&	N/A\\ [0.5ex]
450&	40&	7.233939E-04&	1.060&	1.759571E-04&	1.667\\ [0.5ex]
900&	160&	1.958968E-04&	1.885&	4.733738E-05&	1.894\\ [0.5ex]
1800&	640&	5.243753E-05&	1.901&	1.271703E-05&	1.896\\ [0.5ex]
3600&   2560&   1.374207E-05&  1.932&   3.405083E-06&    1.901\\ [0.5ex]
\hline 
\hline
\end{tabular}
\medskip
\caption{$E_\infty$ and $E_2$ of the Cubic Spline
approximation for pricing of an American put under the Kou Jump-diffusion model are presented. $N$ is the number of the interpolation points. $M_0$ is the number of the time steps. $\hat{x_i}=\log S_i $ is any evaluation points of a range of $S$ from 0.05 to 2 and the total numbers are 1950. T is the Time-to-maturity.
The parameters are: $r = 0,$ $q = 0,$ $\sigma = 0.2,$ $\alpha_1=3,$ $\alpha_2 = 2,$ $\lambda = 0.2,$ $p = 0.5,$ $K=1$ and $T = 0.2.$
The parameters are taken from \citep{Alm_Oos:2005}. The order of convergence is 2 in space and 1 in time.
} 
\label{Chptr4:table:Amer_cubicKou1}
\end{table}

\begin{table}[htbp]
\centering 
\begin{tabular}{c c c c c c} 
\hline
\hline
N & $M_0$ & $E_\infty(\hat{x}_i,T)$ & $R_\infty$ & $E_2(\hat{x}_i,T)$ & $R_2$ \\ [0.5ex] 
\hline
\hline
225&	10&	1.933354E-03&	N/A&	8.983577E-04&	N/A\\ [0.5ex]
450&	40&	8.487095E-04&	1.188&	2.783005E-04&	1.691\\ [0.5ex]
900&	160&	2.497213E-04&	1.765&	7.257535E-05&	1.939\\ [0.5ex]
1800&	640&	6.843085E-05&	1.868&	1.933309E-05&	1.908\\ [0.5ex]
3600&	2560&	1.827216E-05&	1.905&	5.119491E-06&	1.917\\ [0.5ex]
\hline 
\hline
\end{tabular}
\medskip
\caption{$E_\infty$  and $E_2$ of the Cubic Spline
approximation for pricing of an American put under the Kou Jump-diffusion model are presented. $N$ is the number of the interpolation points. $M_0$ is the number of the time steps. $\hat{x_i}=\log S_i $ is any evaluation points of a range of $S$ from 0.05 to 2 and the total numbers are 1950. T is the Time-to-maturity.
The parameters are: $r = 0.05,$ $q = 0,$ $\sigma = 0.15,$ $\alpha_1=3.0465,$ $\alpha_2 = 3.0465,$ $\lambda = 0.1,$ $p = 0.3445,$ $K=1$ and $T = 0.25.$
The parameters are taken from \citep{Carr_Ani_JumpDiff:2007}. The order of convergence is 2 in space and 1 in time.
} 
\label{Chptr4:table:Amer_cubicKou2}
\end{table}
\begin{table}[htbp]
\centering
\begin{tabular}{c c c c c c} 
\hline
\hline
N & $M_0$ & $E_\infty(\hat{x}_i,T)$ & $R_\infty$ & $E_2(\hat{x}_i,T)$ & $R_2$ \\ [0.5ex] 
\hline
\hline
225&	10&	3.839148E-03&	N/A  &	1.095217E-03&	N/A     \\ [0.5ex]
450&	40&	9.616353E-04&	1.997&	2.458977E-04&	2.155   \\ [0.5ex]
900&	160&	2.405238E-04&	1.999&	6.111403E-05&	2.008\\ [0.5ex]
1800&	640&	6.013812E-05&	2.000&	1.508359E-05&	2.019\\ [0.5ex]
3600&	2560&	1.490999E-05&	2.012&	3.768285E-06&	2.001\\ [0.5ex]
\hline 
\hline
\end{tabular}
\medskip
\caption{$E_\infty$ and $E_2$ of the Cubic Spline
approximation for pricing of an American put under the Kou Jump-diffusion model are presented. $N$ is the number of the interpolation points. $M_0$ is the number of the time steps. $\hat{x_i}=\log S_i $ is any evaluation points of a range of $S$ from 0.05 to 2 and the total numbers are 1950. T is the Time-to-maturity.
The parameters are: $r = 0.04,$ $q = 0.03,$ $\sigma = 1,$ $\alpha_1=4,$ $\alpha_2 = 4,$ $\lambda = 0.3,$ $p = 0.6,$ $K=1$ and $T = 1,$ whereas the parameter $\sigma=1$ is selected to stress our numerical algorithm.
The order of convergence is 2 in space and 1 in time.
} 
\label{Chptr4:table:Amer_cubicKou3}
\end{table}

\begin{table}[htbp]
\centering
\begin{tabular}{c c c c c c} 
\hline
\hline
N & $M_0$ & $E_\infty(\hat{x}_i,T)$ & $R_\infty$ & $E_2(\hat{x}_i,T)$ & $R_2$ \\ [0.5ex] 
\hline
\hline
3600&	40&	    1.029993E-03&	N/A  &	3.556438E-04&	N/A  \\ [0.5ex]
3600&	160&	2.974325E-04&	0.896&	8.866477E-05&	1.002\\ [0.5ex]
3600&	640&	8.273457E-05&	0.923&	2.147058E-05&	1.023\\ [0.5ex]
3600&	2560&	2.123568E-05&	0.982&	5.345367E-06&	1.003\\ [0.5ex]
\hline 
\hline
3600&	40&	    1.523857E-03&	N/A&    3.755414E-04&	N/A\\ [0.5ex]
3600&	160&	4.561966E-04&	0.870&	1.108779E-04&	0.880\\ [0.5ex]
3600&	640&	1.253234E-04&	0.932&	2.857790E-05&	0.978\\ [0.5ex]
3600&	2560&	3.230111E-05&	0.978&	7.134578E-06&	1.001\\ [0.5ex]
\hline 
\hline
\end{tabular}
\medskip
\caption{$E_\infty$ and $E_2$ of the Cubic Spline
approximation for pricing of an American put under the Merton Jump-diffusion model are presented. $N$ is the number of the interpolation points. $M_0$ is the number of the time steps. $\hat{x_i}=\log S_i $ is any evaluation points of a range of $S$ from 0.05 to 2 and the total numbers are 1950. Top: The input parameters are provided in the caption to Table \ref{Chptr4:table:Amer_cubicMerton1}. Bottom: The input parameters are provided in the caption to Table \ref{Chptr4:table:Amer_cubicMerton2}.
The order of convergence is 1 in time.
} 
\label{Chptr4:table:Amer_cubicMJ_timstep}
\end{table}
\begin{table}[htbp]
\centering
\begin{tabular}{c c c c c c} 
\hline
\hline
N & $M_0$ & $E_\infty(\hat{x}_i,T)$ & $R_\infty$ & $E_2(\hat{x}_i,T)$ & $R_2$ \\ [0.5ex] 
\hline
\hline
3600&	40&	    1.687267E-03&	N/A  &	1.878158E-04&	N/A \\ [0.5ex]
3600&	160&	4.913041E-04&	0.890&	5.514562E-05&	0.884\\ [0.5ex]
3600&	640&	1.366624E-04&	0.923&	1.590242E-05&	0.897\\ [0.5ex]
3600&	2560&	3.478690E-05&	0.987&	3.768285E-06&	0.923\\ [0.5ex]
\hline 
\hline
3600&	40&	    8.987023E-04&	N/A  &	2.671173E-04&	N/A  \\ [0.5ex]
3600&	160&	3.151073E-04&	0.756&	7.544376E-05&	0.912\\ [0.5ex]
3600&	640&	9.355218E-05&	0.876&	2.184649E-05&	0.894\\ [0.5ex]
3600&	2560&	2.675431E-05&	0.903&	6.204563E-06&	0.908\\ [0.5ex]
\hline 
\hline
\end{tabular}
\medskip
\caption{$E_\infty$ and $E_2$ of the Cubic Spline
approximation for pricing of an American put under the Kou Jump-diffusion model are presented. $N$ is the number of the interpolation points. $M_0$ is the number of the time steps. $\hat{x_i}=\log S_i $ is any evaluation points of a range of $S$ from 0.05 to 2 and the total numbers are 1950. Top: The input parameters are provided in the caption to Table \ref{Chptr4:table:Amer_cubicKou1}. Bottom: The input parameters are provided in the caption to Table \ref{Chptr4:table:Amer_cubicKou2}.
The order of convergence is 1 in time.
} 
\label{Chptr4:table:Amer_cubicKou_timstep}
\end{table}

\section{\bf Conclusion}\label{Chptr4:section:conclusions}
We have implemented an RBF method to solve the PIDE boundary value problem for pricing American put and European call/put options on a dividend-paying stock in the Merton and Kou Jump-diffusion market. By using the numerical scheme of Briani \textsl{et al.}, we find out a finite computational range of our global integral. Our results suggest that the Cubic Spline approximation scheme can achieve second-order convergence in both spatial and time variables (due to the second order time-stepping scheme, BDFs of order 2) when it is used to compute European call/put options. Moreover, the results also show that our approximation solution can get second-order convergence in spatial variables and first-order convergence in time variables when the approximation scheme is used to compute American put options. Beside this, we compare our RBF-approximation method against FDM and FEM. Our results suggest that one can achieve a high accuracy by implementing our meshless scheme. Moreover, in terms of meshless interpolation methods, we use cubic spline as a basis function rather than MQ. This basis function can avoid the open question of choosing an optimal shape parameter of MQ. Beside this, by using factorisation of the Cubic Spline, we can avoid inverting an ill-conditioned cubic spline interpolant directly. Finally, throughout the analysis of both $\Delta$ and $\Gamma$, our RBF-approximation method can also avoid the oscillation problem around the strike $K$ in both American and European cases.

At this stage of development, the Cubic Spline approximation scheme is first order in time for American put options although a second order time-stepping scheme, BDFs of order 2 is implemented. We are investigating various approaches to improve the Cubic Spline approximation for time variables and will treat them in a future paper. Our Method extends in principle to pure jump L\'evy type models for the underlying stocks, like the Variance Gamma (VG) model or the CGMY model.
\clearpage
\appendix
\section{A Finite Computational Range in the Jump-diffusion Model}\label{AppendixA}
In the Merton Model suppose in a domain $\Omega\in\mathbb{R}$ European option price $u(x,\tau)$ satisfies Lipchitz inequality such that
$$|u(x_1,\tau)-u(x_2,\tau)|\leq L|x_1-x_2|,\,\, \forall\, x_1,x_2\in \Omega.$$
Then we choose a parameter $\epsilon>0$ and select the bounded intervals $[y_{-\epsilon}, y_\epsilon]$ as the set of all points $y$ that verify
$$k(y)=\frac{1}{\sqrt{2\pi}\sigma_J}e^{-\frac{(y-\mu_J)^2}{2\sigma_J^2}}\geq\epsilon.$$
Because of the symmetry of $k(y)$ we set $y_{-\epsilon}=-y_\epsilon$. Then the truncation of the integral domain giving an error to approximation of the problem can be estimated by
\begin{subequations}
\begin{eqnarray}
&\,&\left| \int_{-\infty}^\infty (u(x+y)-u(x)) k(y)\,\mathrm{d}y -\int_{-y_\epsilon}^{y_\epsilon} (u(x+y)-u(x)) k(y)\,\mathrm{d}y\right|\nonumber \\
&\leq&L\left| \int_{-\infty}^\infty (x+y - x) k(y)\,\mathrm{d}y -\int_{-y_\epsilon}^{y_\epsilon} (x+y -x) k(y)\,\mathrm{d}y\right| \\
&\leq&L \left(\int_{-\infty}^{-y_\epsilon}|y| k(y)\,\mathrm{d}y+\int_{y_\epsilon}^\infty |y| k(y)\,\mathrm{d}y
\right)\\
&=&2\int_{y_\epsilon}^{\infty}y\frac{1}{\sqrt{2\pi}\sigma_J}\exp(-\frac{(y-\mu_J)^2}{2\sigma_J^2}) \,\mathrm{d}y\\
&=&2\int_{y_\epsilon-\mu_J}^{\infty}(y+\mu_J)\frac{1}{\sqrt{2\pi}\sigma_J}\exp(-\frac{y^2}{2\sigma_J^2}) \,\mathrm{d}y\\
&=&2\int_{y_\epsilon-\mu_J}^{\infty}(y+\mu_J)\frac{1}{\sqrt{2\pi}\sigma_J}\exp(-\frac{y^2}{2\sigma_J^2}) \,\mathrm{d}y\\
&\leq&2\int_{y_\epsilon-\mu_J}^{\infty}(y+y)\frac{1}{\sqrt{2\pi}\sigma_J}\exp(-\frac{y^2}{2\sigma_J^2}) \,\mathrm{d}y\\
&=&\frac{4\sigma_J}{\sqrt{2\pi}}\exp({-\frac{(y_\epsilon-\mu_J)^2}{2\sigma_J^2}})\label{AppendixA:Merton:1} \\
&=&2\sigma_J^2\epsilon\label{AppendixA:Merton:2}
\end{eqnarray}
\end{subequations}
Hence by using (\ref{AppendixA:Merton:1}) and (\ref{AppendixA:Merton:2}),
\begin{eqnarray}
y_\epsilon=\sqrt{-2\sigma_J^2\log(\epsilon\sigma_J\sqrt{2\pi}/2)}+\mu_J
\end{eqnarray}
We use the aforementioned arguments to find the finite computational range $[y_{-\epsilon}, y_{\epsilon}]$ in the Kou model. We carry out the reasoning for the positive semi-axis (the reasoning goes
similarly for the negative semi-axis) and set $k(y)=p\alpha_1e^{-\alpha_1y}$ for $y\geq0$ \big($(1-p)\alpha_2e^{\alpha_2x}$ for $y<0$\big). Then, $y_\epsilon$ can be found out by the following equations:
\begin{subequations}
\begin{eqnarray}
&\,&\left| \int_{0}^\infty (u(x+y)-u(x)) \lambda f(y)\,\mathrm{d}y -\int_{0}^{y_\epsilon} (u(x+y)-u(x)) \lambda f(y)\,\mathrm{d}y\right|\nonumber \\
&\leq&L\left| \int_{0}^\infty (x+y - x) \lambda f(y)\,\mathrm{d}y -\int_{0}^{y_\epsilon} (x+y -x) \lambda f(y)\,\mathrm{d}y\right| \\
&\leq&L \int_{y_\epsilon}^\infty |y| f(y)\,\mathrm{d}y \\
&=&\int_{y_\epsilon}^\infty |y| p\alpha_1e^{-\alpha_1y}\,\mathrm{d}y \\
&=&p\alpha_1e^{-y_\epsilon\alpha_1}\left(\frac{1}{\alpha_1^2}+\frac{y_\epsilon}{\alpha_1}\right) \\ &\ &\hbox{\citep[equation 3.351]{Gra_Ryz:1994}}\nonumber\\
&=&\frac{p}{\alpha_1}e^{-y_\epsilon\alpha_1}(1+y_\epsilon\alpha_1)\\
&\leq&\frac{p}{\alpha_1}e^{-y_\epsilon\alpha_1}\alpha_1e^{y_\epsilon}\\
&=&pe^{y_{\epsilon}(1-\alpha_1)}\\
&=&\epsilon,
\end{eqnarray}
\end{subequations}
as a result,
\begin{eqnarray}
y_\epsilon=\log(\epsilon/p)/(1-\alpha_1).
\end{eqnarray}
Similar arguments can be applied to $y<0$, so
\begin{eqnarray}
y_{-\epsilon}=-\log\big(\epsilon/(1-p)\big)/(1-\alpha_2).
\end{eqnarray}
\bibliographystyle{rAMF}
\bibliography{JumpDiff}
\vspace{36pt}
\end{document}